\documentclass[%
 reprint,
superscriptaddress,
 amsmath,amssymb,
pra,
]{revtex4-2}
\usepackage{float}
\usepackage{lipsum}
\usepackage{caption}
\usepackage{subcaption}
\usepackage{gensymb}
\usepackage[export]{adjustbox}
\usepackage{xcolor}
\usepackage{graphicx}
\usepackage{graphicx}
\usepackage{dcolumn}
\usepackage{fontenc}
\usepackage{bm}
\usepackage{multirow}
\DeclareUnicodeCharacter{2212}{-}

\begin{document}
\title{Mechanical response of butylamine based Ruddlesden Popper organic-inorganic lead halide perovskites}

\author{Yashika Gupta}
\thanks{These two authors contributed equally}
 \affiliation{Department of Physics, Indian Institute of Technology Bombay}
\author{Sudharm Rathore}
\thanks{These two authors contributed equally}
 \affiliation{Department of Metallurgical Engineering and Materials Science, Indian Institute of Technology Bombay}
 \author{Aparna Singh}
 \email{aparna$_$s@iitb.ac.in}
 \affiliation{Department of Metallurgical Engineering and Materials Science, Indian Institute of Technology Bombay}
\author{Anshuman Kumar}
\email{anshuman.kumar@iitb.ac.in}
 \affiliation{Department of Physics, Indian Institute of Technology Bombay}

\begin{abstract}
2D Ruddlesden Popper perovskites have been extensively studied for their exceptional optical and electronic characteristics while only a few studies have shed light on their mechanical properties. The existing literature mainly discusses the mechanical strength of single crystal perovskites, however a systematic study towards structure tunability of 2D perovskite thin films is still missing. In this study, we report the effect of number of inorganic layers `n' on elastic modulus of Butylammonium based 2D, quasi-2D perovskites and 3D perovskite using nanoindentation technique. The experimental results have also been substantiated using first principle density functional theory calculations. Understanding the mechanical behaviour of 2D Ruddlesden Popper perovskites thin films in comparison with conventional 3D perovskite offers intriguing insights into the atomic layer dependent properties and paves the path for next generation mechanically durable novel devices.

\end{abstract}

\maketitle
\section{Introduction}
\begin{figure*}[h]
    \centering
    \includegraphics[width=1.0\linewidth]{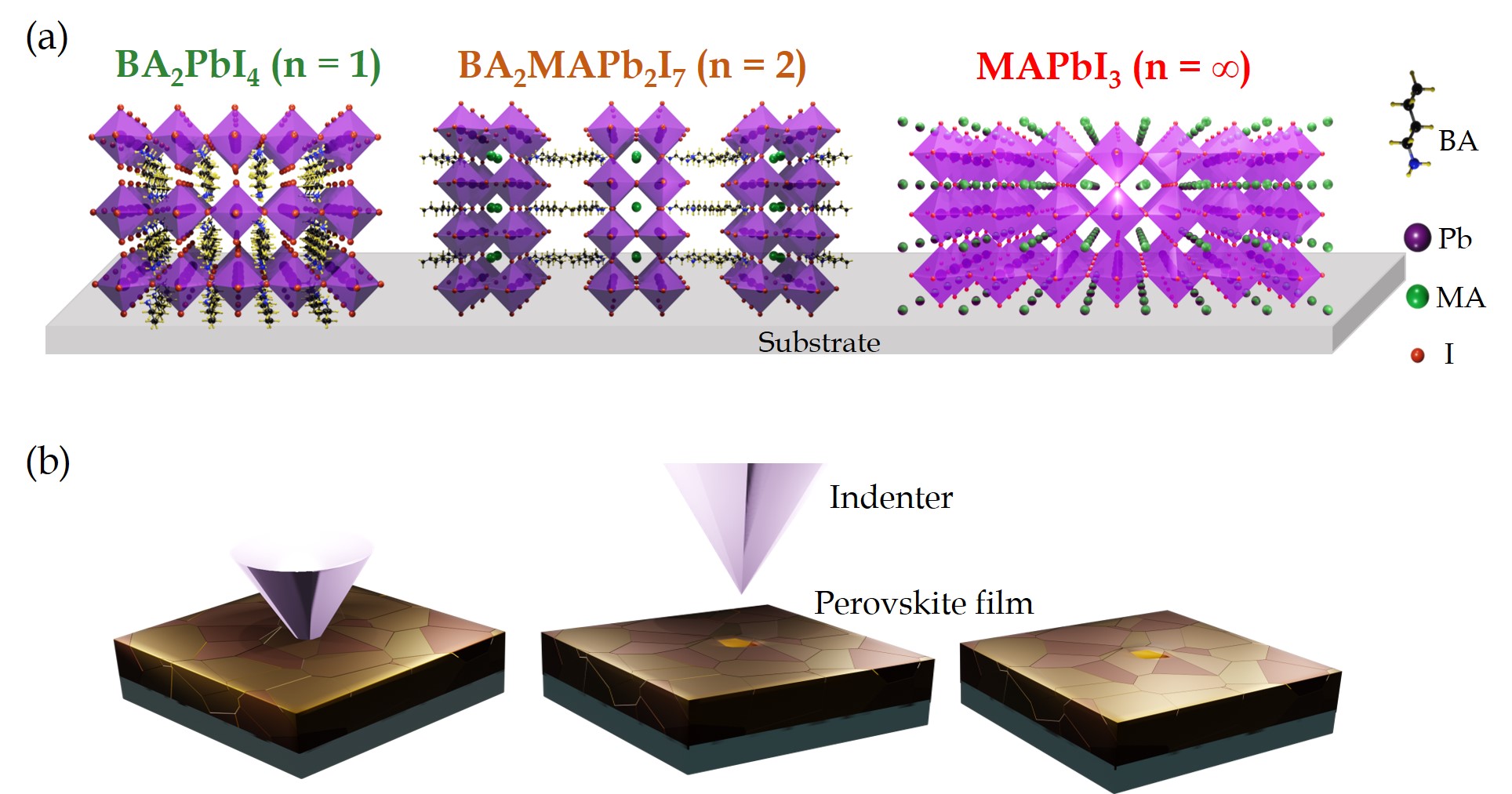}
    \caption{Schematic illustration of (a) arrangement of octahedra layers in 2D perovskite ${\rm BA_2PbI_4}$ (n1) , quasi-2D perovskite ${\rm BA_2MAPb_2I_7}$ (n2) and 3D perovskite ${\rm MAPbI_3}$ thin films under study with respect to the substrate.(b) nanoindentation measurement for a perovskite thin film sample.}
    \label{fig:Figure_1}
\end{figure*}

Metal halide organic-inorganic perovskites, having characteristic formula, ${\rm ABX_3}$, where A = ${\rm {CH_3NH_3}^+}$ (MA), ${\rm {HC(NH_2)_2}^+}$ (FA) etc; B = ${\rm {Pb_2}^+}$ ; X = ${\rm Cl^−}$, ${\rm Br^−}$, ${\rm I^−}$, possess favourable properties like high absorption coefficient (${\rm 10^4}$-${\rm 10^5}$~${\rm cm^{-1}}$)~\cite{Maes2018, Park2015}, low exciton binding energy (10-60~meV)~\cite{Ruf2019}, tunable direct band-gap (2.3-1.5~eV)~\cite{Wang2016, Dong2017}, long charge diffusion length (${\rm >10~\mu m}$)~\cite{Herz2017, Dittrich2016} and defect tolerance~\cite{Chen2018}. These materials have revolutionized thin film technology, showing great potential in applications ranging from photovoltaics to light emitting devices to transistors to synaptic devices. However, poor stability of these perovskites towards ambient light, moisture and temperature raises a big question towards the commercialization of this material class. To remedy this situation, researchers have come up with a solution of ${\rm ``slicing"}$ the conventional perovskite structures (${\rm A BX_3}$) into a 2D structure of the form ${\rm R_2 A_{n-1} Pb_n X_{3n+1}}$. This change in dimensionality is brought by replacing or mixing the `A' cation by a larger monovalent organic cation `R' (Butylamine (BA), Phenylethylamine (PEA) etc.) giving perovskite a layered structure as shown in Figure~\ref{fig:Figure_1}(a). These 2D perovskites also known as Ruddlesden Popper perovskites, are classified by the number of inorganic layers `n' in between the large organic cation. Perovskite with $n=1$ contains horizontal inorganic layers stacked vertically together by Van der Waal forces due to interaction between the organic cations of these layers, showing a complete two-dimensional behaviour. As the number of layers increases,the structure goes from 2D to quasi-2D to finally a 3D structure (see Figure~\ref{fig:Figure_1}(a)). These two-dimensional perovskites not only exhibit exceptional optoelectronic properties like 3D perovskites, they also hold the potential of great structural tunability, arising due to its layered structure.The Van der Waal forces holding organic and inorganic layers together, opens great avenues for flexible applications like printed solar cells, foldable sensors, wearable devices etc.

Building efficient and reliable flexible perovskite devices requires a thorough understanding of interaction mechanisms of inorganic and organic layers in these materials' thin films. A few mechanical studies, both theoretical and experimental have been conducted in the past to explore their potential for wearable electronics. Tu et. al. have reported in-plane and out of plane mechanical properties of a series of 2D Lead-Halide perovskites single crystals~\cite{Tu2018_i, Tu2018}. They reported a decrease in in-plane breaking strength and elastic modulus as one goes from monolayer to trilayer for pure BA-based 2D crystals. While out of plane inspection showed an increased stiffness with increasing $n$ and shorter chain ${\rm `R'}$ cations. Spanopoulos et. al. carried out a similar study on three different 2D perovskite crystals namely, pentlyamine-, hexylamine-, butylamine- lead iodide, each having four inorganic slabs in between organic layers~\cite{Spanopoulos2019}. They also observed an increase in elastic modulus and hardness values for smaller carbon-chained perovskites. Recently, Reyes-Martinez et. al. have published an extensive experimental and theoretical study on out of plane and in-plane mechanical properties of phenylethylammonium methylammonium lead iodide single crystals ${\rm PEA_2MA_{n-1}Pb_n I_{n−1}}$ ~\cite{ReyesMartinez2020} and reported a non-monotonic variation in elastic moduli for these crystals with varying number of inorganic layers ($n$). Li et. al. have presented the variation in shear modulus for 2D ${\rm (C_4H_9NH_3)_2Pb Br_4}$ perovskite crystal with decreasing number of atomic layers along (001) crystal plane in their work~\cite{Li2019}. Gao et. al. reported mechanical properties of benzylammonium lead chloride single crystals~\cite{Gao2020} along (100), (001), and (110) crystal planes. Recently Singh et. al. reported the effect on elastic modulus of the 2D-3D mixed perovskite thin films by varying the  fraction of 5-aminovaleric acid cation based 2D perovskite within the system~\cite{Rathore2021}. While all these studies provide an insight into the structural dynamics of these materials, the 2D perovskite literature still lacks a systematic mechanical study, specially on thin films, which is highly relevant from the technological point of view.

In this manuscript, we report variation of elastic properties of BA-based 2D perovskite thin films with increasing number of inorganic slabs,namely $n=1,2$ and ${\rm \infty}$ (labelled as n1, n2 and 3D respectively henceforth) measured using nanoindentation technique as shown in Figure~\ref{fig:Figure_1}(b).  Using DFT calculations, we show that the experimentally observed mechanical response shows the expected behaviour as a function of $n$. This study can serve as a building foundation to establish designing principles for the application of 2D perovskites in bendable technology.

\section{Experimental}

\subsection{Materials and Synthesis}
Perovskite thin film samples were prepared on glass substrates ($1 \times 1$ cm) using a solution based method inside a nitrogen-filled glove box. The starting materials BAI (Butylammonium Iodide), MAI (Methylammonium Iodide), ${\rm PbI_2}$ (Lead Iodide) and DMF (Dimethylformamide) were procured from Sigma-Aldrich. For making 2D Ruddlesden Popper perovskite thin film (n = 1), 1~M ${\rm PbI_2}$ with 1~M BAI were mixed together in DMF at ${\rm 70^{\circ}C}$ to obtain pure 2D perovskite solution (${\rm BA_2PbI_4}$). For quasi-2D perovskite thin film (n = 2), 2D perovskite solution was mixed  in equal proportion (1:1) with another solution prepared by mixing 1~M ${\rm PbI_2}$ and 1~M MAI in DMF at ${\rm 70^{\circ}C}$. The obtained solutions were spin coated on a pre-heated glass substrates (${120^{\circ}}$C) at 1000~rpm for 10~s and then at 4000~rpm for 45~s. After coating, substrates were immediately transferred to a hot plate for annealing at ${120^{\circ}}$C for 20 minutes. 

For drawing comparison with standard 3D perovskite, we also deposited ${\rm MAPbI_3}$ thin films by using two-step coating approach. In this technique, first ${\rm 1.5~M}$ ${\rm PbI_2}$ layer was spin coated on a glass substrate at 3000~rpm for 30~s and annealed on a hot plate at ${60^{\circ}}$C for 5 minutes. The molarity of the solution was chosen  to get the film thickness of same range as that of 2D perovskite thin films. The perovskite phase was subsequently obtained by spin coating MAI solution on the as-developed ${\rm PbI_2}$ thin film at 5000~rpm for 30~s and then annealing at ${60^{\circ}}$C for 15 minutes. The whole procedure of depositing perovskite thin films (both 2D and 3D) was done inside a nitrogen glovebox.

\subsection{Optical and Structural Characterization}
The thickness and the roughness of the as-deposited perovskite thin films were measured using Dektak surface profilometer. All the films were found to have a thickness of ${\rm \sim1~\mu m}$ with surface roughness less than 30~nm. The X-ray diffraction was done using Bruker DA advanced XRD operated under the standard Bragg's ${\rm \theta-2\theta}$ mode with Cu-${\rm K\alpha}$ radiation at a scanning rate of ${\rm 0.1^{\circ}/s}$. The steady state absorption spectra of the films were studied with a Perkin Elmer lambda 950 UV-Vis spectrometer. The emission spectra of the films were recorded by exciting the samples with 488~nm wavelength using Horiba Scientific ${\rm Fluoromax-4}$ spectrofluorometer. 

\subsection{Nanoindentation}
The modulus of elasticity of the samples was measured using nanoindentation technique at room temperature, using a TI Premier Triboindenter, HYSITRON Inc. (USA). The nanoindentation test was conducted in depth-controlled mode, using a three-sided, sharp pyramidal Berkovich diamond tip (tip radius 100 nm) indenter which is aligned normal to the film thickness. Load-displacement data obtained from the test was used to calculate the elastic modulus of the as-fabricated perovskite films according to the Oliver and Pharr method~\cite{Oliver1992, Pharr1992}. Ten indents were performed on each sample with the indentation depth of 70~nm for each indent and the average of the ten indents reading has been reported. The Young’s modulus (E) value of 1141~GPa and Poisson’s ratio (${\rm \nu}$) values of 0.07 for the diamond tip~\cite{Oliver2004} and ${\rm \nu = 0.3}$ for hybrid organic-inorganic perovskites~\cite{Sun2015,Feng2014, Rakita2015} were used for the calculation of elastic modulus for 2D and 3D perovskites.

\section{Density Functional Theory Calculations}
Density functional theory (DFT), as implemented in the ${\rm thermo\_pw}$ package~\cite{DalCorso2016}, a FORTRAN user interface that uses the routines of Quantum-Espresso~\cite{Giannozzi2009, Giannozzi2017}, was used for calculating the elastic modulus for n1, n2 and 3D perovskite samples. The analyses were performed within the plane-wave/pseudopotential formalism with a kinetic energy cutoff of 45 Ry to represent the wave function and a cut of 450 Ry for the electronic density. Projector augmented wave (PAW) scalar relativistic pseudopotentials were used for all the elements that are carbon, hydrogen, nitrogen, iodine and lead with Perdew-Burke-Ernzerhof (PBE) exchange-correlation functional~\cite{Perdew1998}. The Brillouin zone was sampled with Monkhorst-Pack scheme of k point grids of ${\rm 4 \times 4 \times 1}$, ${\rm 5 \times 5 \times 1}$, ${\rm 4 \times 4 \times 2}$ for n1, n2 and 3D perovskite structures. Six different deformation modes, three pertaining to non-shearing mode and three pertaining to shearing mode, are applied to the relaxed structure, with one independent deformation considered each time. For each deformation mode, four different magnitudes of strain, 0.0075, 0.0025, - 0.0025 and - 0.0075 have been used. The stress tensor is calculated for each independent strain after the code relaxes the ions to equilibrium. The elastic constants are calculated from the numerical first derivative of the stress with respect to strain. Elastic modulus has been subsequently calculated from the stiffness tensor and is reported for Voigt, Reuss and Hill averages. In addition to this, other important mechanical parameters that are shear modulus, bulk modulus, Poisson's ratio, Pugh's ratio (B/G)~\cite{Pugh1954}, Vickers hardness ${\rm (H_V = 2{\{k^2G\}}^{0.585} – 3)}$, where k= G/B~\cite{Chen2011}, yield strength ${\rm (\sigma_y = H_V/3)}$~\cite{Zaddach2013} and the universal elastic anisotropic index ${\rm (A^U = 5 G_V/G_R + B_V/B_R – 6 \geq 0)}$~\cite{Ranganathan2008} has also been calculated.

\section{Results and Discussion}

\subsection{Optical Characterization}
\begin{figure*}[h]
    \centering
    \includegraphics[width=1.0\linewidth]{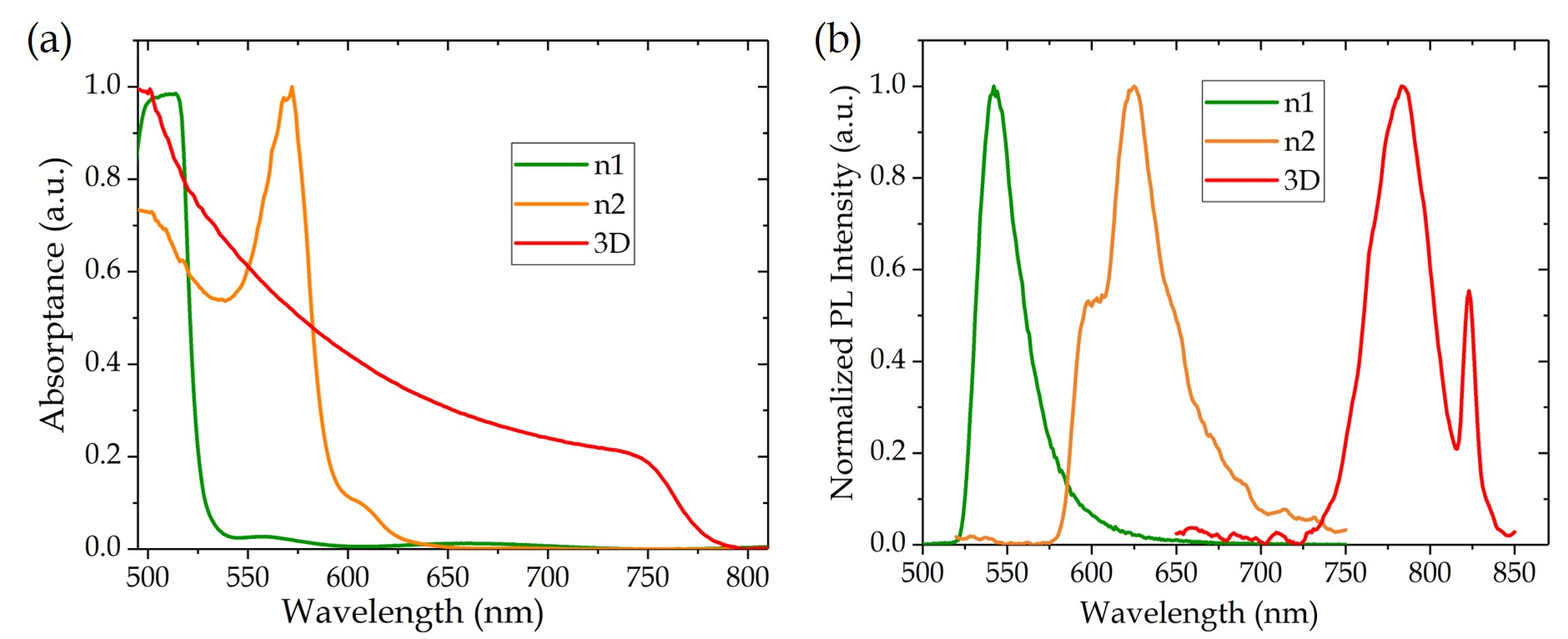}
    \caption{Figure showing the (a) Absorptance and (b) Photoluminescence (PL) spectra for n1, n2 and 3D samples. The spectra show a clear red shift as 'n' increases. }
    \label{fig:Figure_2}
\end{figure*}

The effects of increasing number of inorganic layers ($n$) in our perovskite thin film samples can be easily seen in the absorptance and PL spectra shown in ${\rm Figure~\ref{fig:Figure_2}}$, where the absorption edge and the PL emission peak shows a clear red shift with increasing values of $n$. Sharp excitonic peak at 540~nm for ${\rm BA_2PbI_4}$ sample confirms the 2D nature of the perovskite film with $n=1$, demonstrating its layered quantum well structure. While the broadening of the absorption spectra, covering almost the entire visible range, with absorption edge reaching ${\rm ~760~nm}$ for ${ n=\infty}$ i.e. ${\rm MAPbI_3}$ sample, indicates increasing dimensionality of as-deposited perovskite films from completely  2D to 3D. PL spectra for 3D perovskite shows a prominent PL peak corresponding to the band-gap emission along with a small and narrow shoulder peak. This second peak can be attributed to self-absorption of the emitted photons in the sample~\cite{Schtz2020}.  The absorption edge and PL peak positions for n1, n2 and 3D films agrees well with those reported in literature~\cite{Hwang2018,Schtz2020, Blancon2017}.

\subsection{Structural Characterization}
\begin{figure*}[h]
    \centering
    \includegraphics[width=1\linewidth]{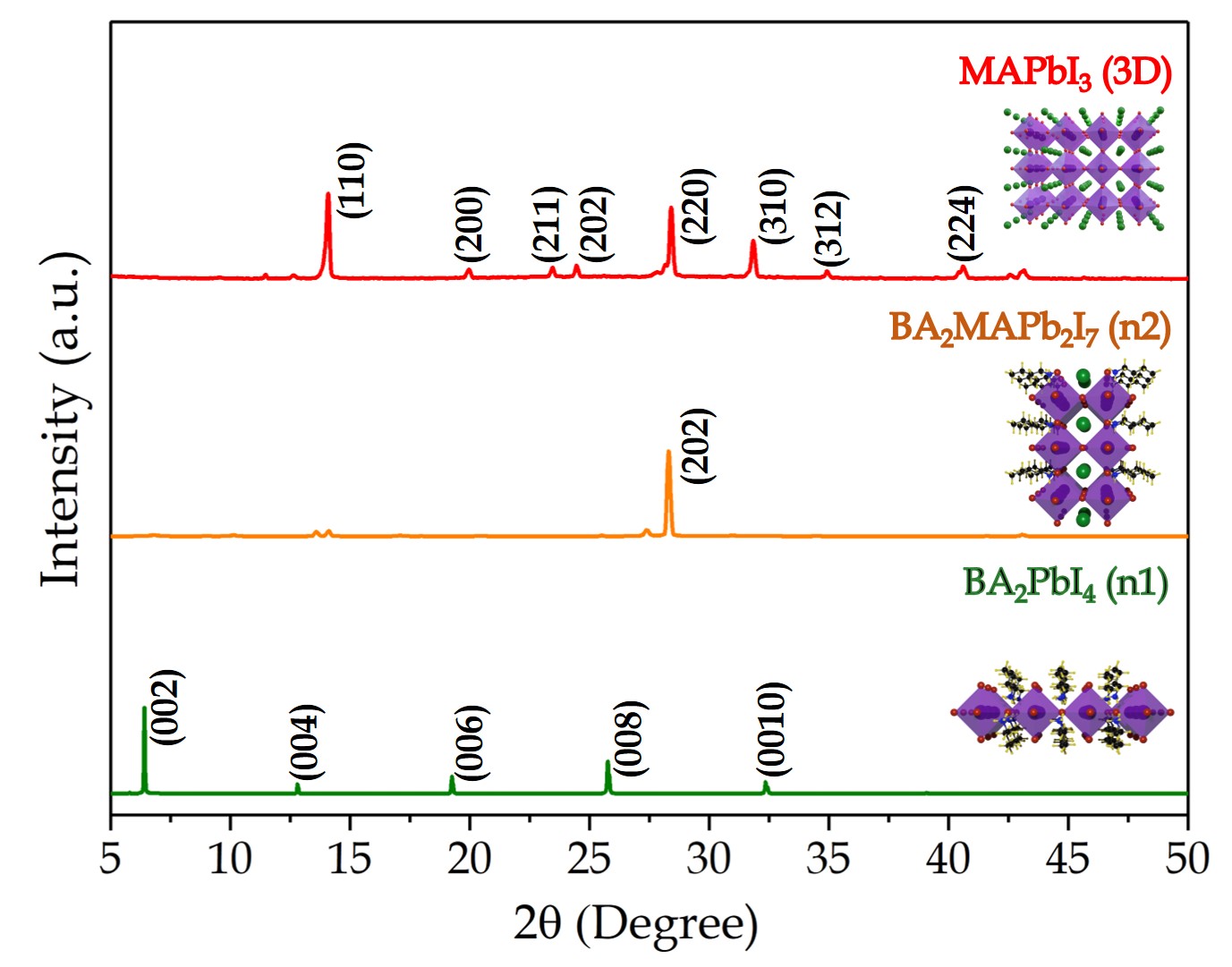}
    \caption{Figure showing X-ray diffractograms for perovskite thin films under study. Inset shows a change in orientation of inorganic slabs with increasing `n'.}
    \label{fig:Figure_3}
\end{figure*}

Due to close relationship between the number of inorganic layers (or $n$) and the structural properties of RP perovskite films, the knowledge of  orientation of inorganic layers in each sample with respect to the underlying layers (substrate in this case) becomes vital for various device applications based on this class of materials. ${\rm Figure~\ref{fig:Figure_3}}$ shows the X-ray diffractograms obtained for the films under study. The phase pure ${\rm BA_2PbI_4}$ (n1) film showed peaks corresponding to (00l) crystal planes~\cite{Cao2015} only at ${\rm \theta = 6.48^\circ}$, ${\rm 12.92^\circ}$, ${\rm 19.37^\circ}$, ${\rm 25.89^\circ}$ and ${\rm 32.39^\circ}$. The presence of these intense, sharp peaks shows  high crystallinity of the sample with the perovskite layers arranged parallel to the substrate with horizontal quantum well structure as depicted in the inset. The absence of these peaks in samples ${n > 1}$ indicated a shift in orientation of crystal planes in these films as the number of inorganic interlayers were increased. Instead the sample ${\rm BA_2MAPb_2I_7}$ with $n = 2$ showed an intense peak at ${\rm 28.33^\circ}$ corresponding to (202) orientation of crystal planes. The peak corresponding to (202) crystallographic plane indicates perfect orthogonal alignment of ${\rm [MA_{n-1}Pb_nI_{3n+1}]^{2-}}$ slabs with the substrate~\cite{Soe2017} as is shown in inset of the Figure~\ref{fig:Figure_3}, suggesting vertical orientation of the perovskite inorganic layers with respect to the underlying substrate~\cite{Venkatesan2018}. The main reason behind this change of orientation is the competition between BA and MA species to form different perovskite phases. BA cations in the precursor solution try to limit the vertical growth as opposed to MA cations~\cite{Qiu2018} and therefore result in completely parallel inorganic slabs for $n=1$ 2D RP films having no MA cations. The above analysis shows that the $n$-value of RP perovskite thin films not only affects their optical properties but also result in a shift from parallel for $n=1$ films to vertical orientation of inorganic layers for films with ${n>1}$. Being an anisotropic class of materials with two components (large spacer organic cation and metal-halide inorganic anion), these films will respond differently to strain/stress, thus making the knowledge of orientation critical for mechanical testing of the films.

As we go from 2D to 3D perovskite films, X-ray diffractogram showed preferred orientation with intense peaks at ${\rm 2\theta = 14.08^\circ, 28.42^\circ, 31.85^\circ}$, these peaks correspond to (110), (220), (310) crystal planes of ${\rm MAPbI_3}$ perovskite with tetragonal crystal structure~\cite{fan2016}. Minor peaks corresponding to other crystal planes were also present (see ${\rm Figure~\ref{fig:Figure_3}}$) indicating good quality phase pure ${\rm MAPbI_3}$ perovskite films.

\subsection{Nanoindentation}
\begin{figure*}[h]
    \centering
    \includegraphics[width=1.0\linewidth]{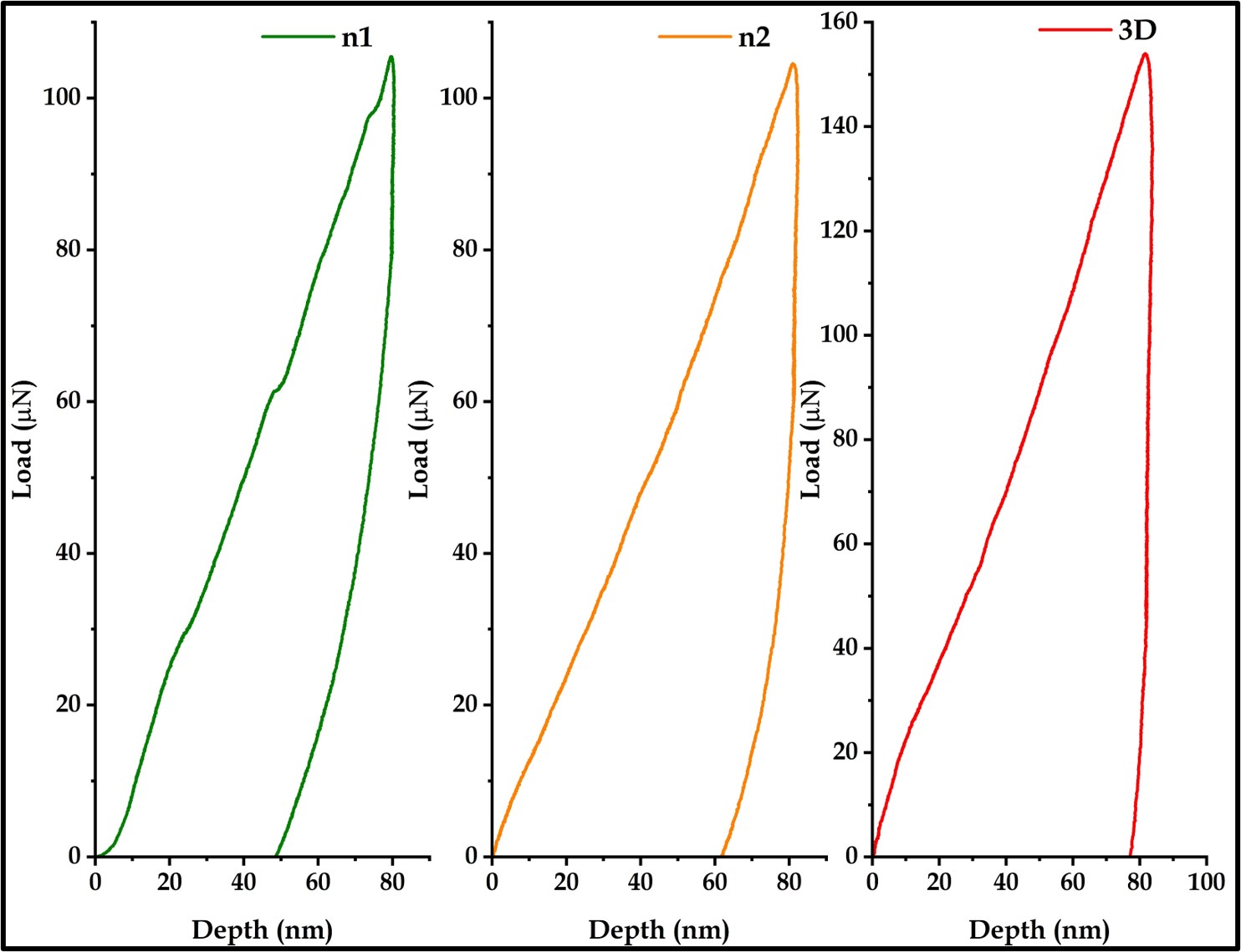}
    \caption{Load vs Depth curve of 2D, quasi-2D and 3D perovskite obtained using nanoindentation.}
    \label{fig:Figure_4}
\end{figure*}

\begin{figure*}[h]
    \centering
    \includegraphics[width=0.75\linewidth]{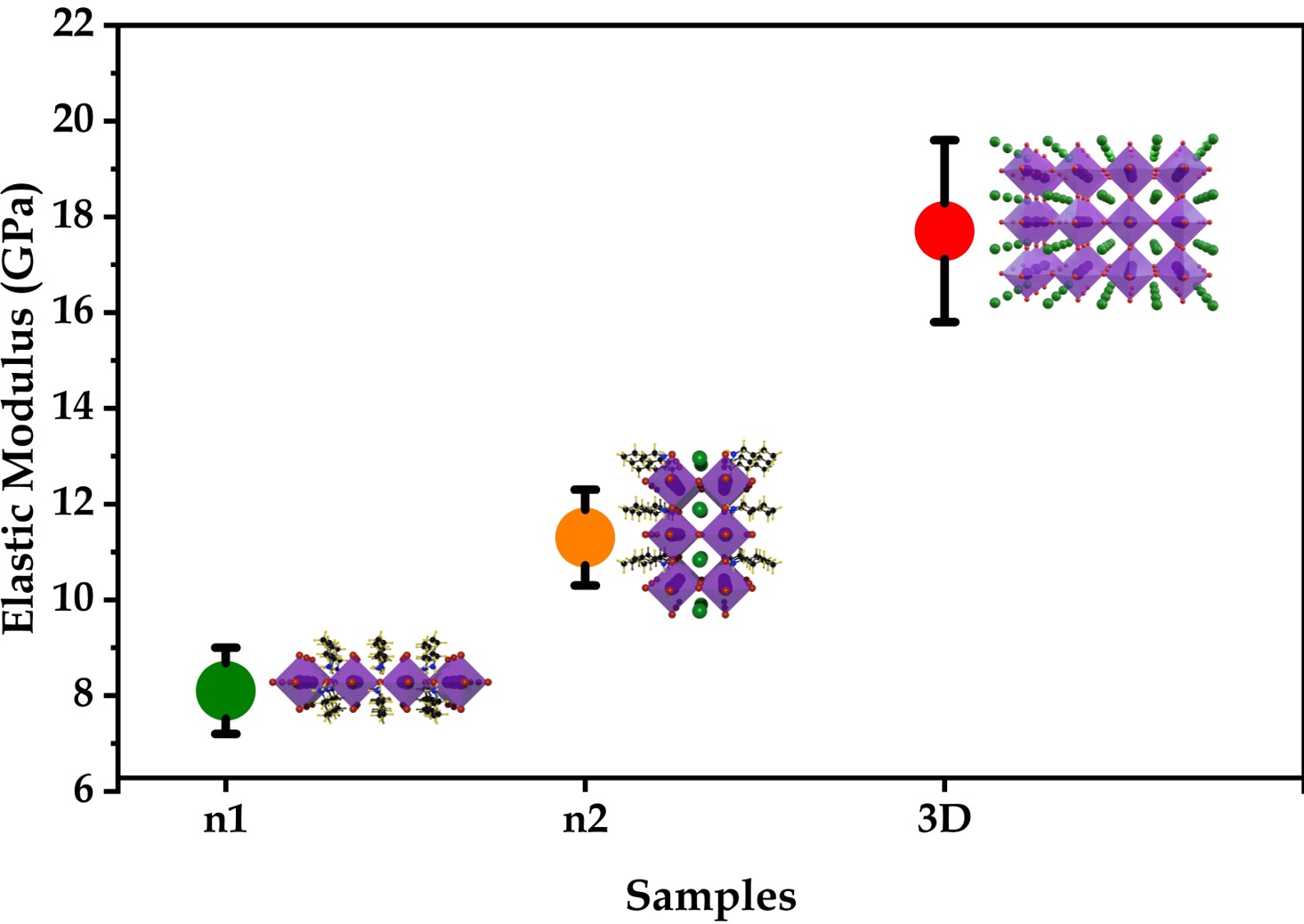}
    \caption{Elastic modulus of 2D, quasi-2D and 3D perovskite thin films on a plain glass substrate measured using nanoindentation.}
    \label{fig:Figure_5}
\end{figure*}

The modulus of elasticity for the Butylamine (BA) based Ruddlesden-Popper (RP) perovskite thin films, with different number of inorganic layers ${n = 1,2}$ and 3D perovskite were measured using nanoindentation. Figure~\ref{fig:Figure_4} represents the typical load vs displacement sample for different samples of BA based 2D (n1), quasi-2D ${\rm (n>1)}$ and 3D perovskite. Standard Oliver and Pharr method have been used to analyze the load vs displacement data~\cite{Pharr1992, Oliver1992}, through which the out of plane elastic modulus of BA based 2D, quasi 2D and 3D perovskite has been calculated. Figure~\ref{fig:Figure_5} shows the elastic modulus calculated for the n1 ${\rm (8.1 \pm 0.9~GPa)}$, n2 ${\rm (11.3 \pm 1.0~GPa)}$ and 3D ${\rm (17.7 \pm 1.9~GPa)}$ perovskite films through nanoindentation. The reported values are the average of the modulus values of ten indents performed on each sample. Thus, from these calculated values of elastic modulus, we can see that the elastic modulus of the 2D, quasi-2D and 3D perovskite samples in our study follows the trend as ${\rm n1<n2<3D}$. This shows that the elastic modulus is least for the n1 sample and increases with an increase in the value of $n$. In other words, we can say that the modulus of the 2D perovskite film keeps on increasing as the structure approaches to that of pure 3D perovskite.\\

The modulus of the 3D hybrid organic-inorganic perovskites depends mainly upon the metal halide that is B−X (B = Pb, Sn; X = Cl, Br, or I) bond strength~\cite{Sun2017, Li2014, Tan2012}. However, in addition to metal halide bond strength, the elastic modulus for the 2D perovskites also depends upon the thickness of organic and inorganic layers and the interfaces between the organic layers~\cite{Tu2018, Tu2020}. Thus, we can see that the modulus of the 2D perovskites samples in our study follows the trend ${\rm n1 < n2 < 3D}$ because of the fact that in the perovskites with higher $n$ number, some inorganic layers get replaced by the soft organic layers in lower $n$ number perovskites. These organic layers interact with each other via weak Van Der Waals forces, which are much weaker forces (an order of magnitude less) as compared to than the strong electrostatic forces present within the inorganic layers. Thus, for the n1 sample with a single inorganic layer sandwiched between the organic layers we get the least modulus as most of the inorganic layers get replaced by the soft organic layers.\\

As seen from the XRD profiles of these samples in Figure~\ref{fig:Figure_3}, the presence of only (h00) peaks shows that for n1 perovskite films, the inorganic octahedra grow parallel to the glass substrate. However, for n2 and 3D samples, the complete absence of (h00) peaks and the presence of (202) and (110) peaks respectively, shows that the inorganic octahedra for these 2D perovskites grow perpendicularly to the substrate. Because of this preferred orientation of thin films to the substrate, the n1 sample is bonded covalently or ionically in-plane and by weak Van der Waal’s forces in out of plane while for n2 sample it's vice-versa~\cite{Cao2015}. Thus, for n1 sample the reason for low elastic modulus can be attributed to combined effect of both the texture and structure of the films.

\subsection{Density Functional Theory Calculations}
We calculated the elastic modulus for n1, n2 and 3D perovskites using DFT. The elastic modulus value (considering Voigt, Reuss and Voigt-Reuss-Hill approximation) is in between 8-9~GPa, 11-13~GPa and 16-19~GPa for n1, n2 and 3D perovskite respectively. The obtained theoretical and experimental values of the elastic modulus are in agreement with each other. In addition to the elastic modulus, other important mechanical parameters like shear modulus, bulk modulus, poisson's ratio, pugh's ratio, hardness, yield strength and universal anisotropic index have also been calculated and reported in Tabel~\ref{tabel:mech}. The Pugh's ratio is an indicatory parameter for the ductility of a material. Any material is considered ductile if the Pugh's ratio is ${\rm >1.75}$ or else it is considered brittle~\cite{Pugh1954}. The calculated Pugh's ratio for n1, n2 and 3D perovskite is ${\rm >1.75}$ and thus, all the three perovskites can be regarded as ductile. The resistance to deformation of a material is represented by hardness as well as yield strength. The hardness value of the material indicates the resistance of the material towards localized surface deformation, whereas the yield strength value suggests how resistant the material is to the plastic deformation~\cite{Chen2011, Zaddach2013}. The hardness and yield strength of both n1 and n2 for all the three types approximations (Voigt, Reuss and Hill) is less than to that of 3D perovskite, and thus it can be said that 2D and quasi-2D perovskite are less resistant to deformation as compared to their 3D counterpart.

\begin{figure*}[h]
    \centering
    \includegraphics[width=1.0\linewidth]{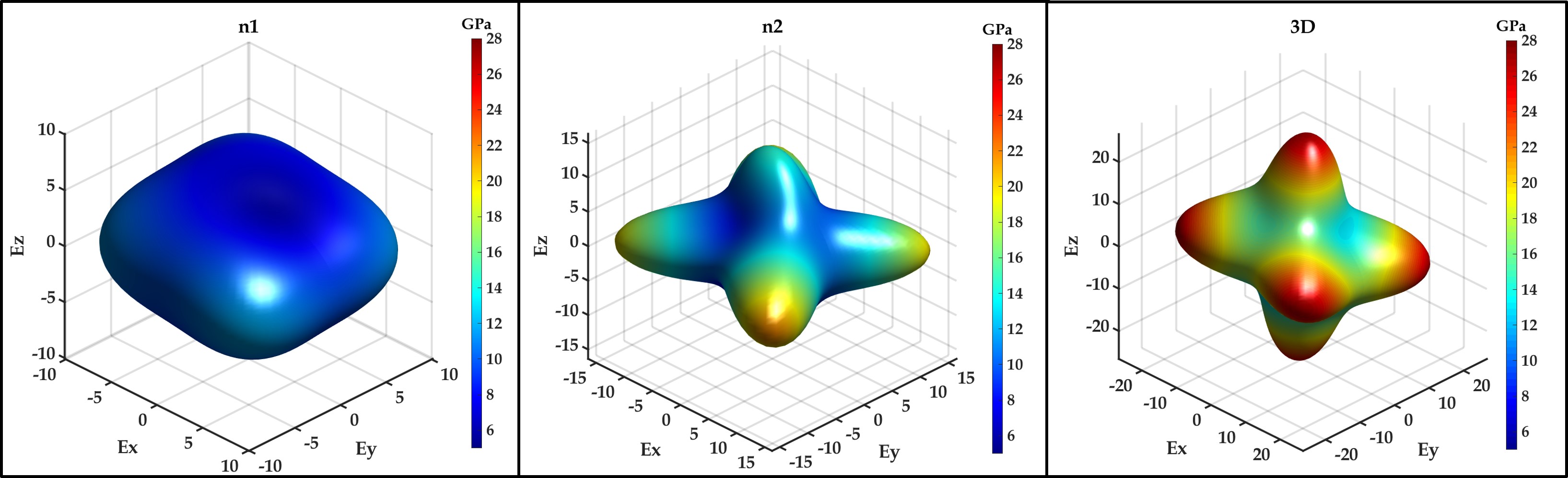}
    \caption{The surface contours of the elastic modulus of n1, n2 and 3D perovskite plotted in 3D space.
    \label{fig:Figure_6}}
\end{figure*}

Elastic anisotropy is another parameter which is considered to be important for material application. It is represented by the universal elastic anisotropic index ${\rm A^U}$. An isotropic material has ${\rm A^U = 0}$, while any non-zero value of ${\rm A^U}$ indicates anisotropy in the material~\cite{Ranganathan2008}. Thus, from the calculated ${\rm A^U}$ values for n1 (${\rm A^U}$ = 0.7), n2 (${\rm A^U}$ = 1.1) and 3D (${\rm A^U}$ = 0.9) perovskites, it can be said that all these materials show anisotropic behavior. To demonstrate the elastic anisotropy of these n1, n2 and 3D perovskites, a three-dimensional (3D) surface contour plot of the elastic modulus has been constructed for each kind of perovskite as shown in Figure~\ref{fig:Figure_6}. It can be seen that the 2D, quasi-2D as well as 3D perovskite compounds, all show a strong elastic anisotropic property.

\begin{table}[]
\caption{Elastic modulus and other mechanical parameters of n1, n2 and 3D perovskite films calculated using the first principle DFT calculations. 
\label{tabel:mech}}
\begin{tabular}{|c|c|c|c|}
\hline
\multirow{2}{*}{\textbf{Sample}}        & \multirow{2}{*}{\textbf{n1}} & \multirow{2}{*}{\textbf{n2}} & \multirow{2}{*}{\textbf{3D}} \\
                                        &                              &                              &                              \\ \hline
\multicolumn{4}{|c|}{\multirow{4}{*}{\textit{\textbf{Voigt approximation}}}}                                                         \\
\multicolumn{4}{|c|}{}                                                                                                               \\
\multicolumn{4}{|c|}{}                                                                                                               \\
\multicolumn{4}{|c|}{}                                                                                                               \\ \hline
Bulk modulus   B (GPa)                  & 8.2                          & 12.4                         & 13.5                         \\ \hline
Elastic   modulus E (GPa)               & 9.2                          & 13.2                         & 19.5                         \\ \hline
Shear modulus   G (GPa)                 & 3.5                          & 5.0                          & 7.8                          \\ \hline
Poisson's   Ratio ${\rm \nu}$                     & 0.31                         & 0.32                         & 0.30                         \\ \hline
Pugh's Ratio   B/G                      & 2.4                          & 2.5                          & 1.7                          \\ \hline
Vickers   Hardness ${\rm H_v}$ (GPa)    & 1.5                          & 1.8                          & 3.5                          \\ \hline
Yield   Strength ${\rm \sigma_y}$ (GPa) & 0.5                          & 0.6                          & 1.2                          \\ \hline
\multicolumn{4}{|c|}{\multirow{4}{*}{\textit{\textbf{Reuss approximation}}}}                                                         \\
\multicolumn{4}{|c|}{}                                                                                                               \\
\multicolumn{4}{|c|}{}                                                                                                               \\
\multicolumn{4}{|c|}{}                                                                                                               \\ \hline
Bulk modulus   B (GPa)                  & 6.5                          & 12.1                         & 13.5                         \\ \hline
Elastic   modulus E (GPa)               & 8.2                          & 11.0                         & 17.5                         \\ \hline
Shear modulus   G (GPa)                 & 3.2                          & 4.1                          & 6.8                          \\ \hline
Poisson's   Ratio ${\rm \nu}$                     & 0.29                         & 0.35                         & 0.30                         \\ \hline
Pugh's Ratio   B/G                      & 2.0                          & 3.0                          & 2.0                          \\ \hline
Vickers   Hardness ${\rm H_V}$ (GPa)    & 1.7                          & 1.3                          & 2.8                          \\ \hline
Yield   Strength ${\rm \sigma_y}$ (GPa) & 0.6                          & 0.4                          & 0.9                          \\ \hline
\multicolumn{4}{|c|}{\multirow{4}{*}{\textit{\textbf{Voigt - Reuss - Hill Average}}}}                                                \\
\multicolumn{4}{|c|}{}                                                                                                               \\
\multicolumn{4}{|c|}{}                                                                                                               \\
\multicolumn{4}{|c|}{}                                                                                                               \\ \hline
Bulk modulus   B (GPa)                  & 7.3                          & 12.3                         & 13.5                         \\ \hline
Elastic   modulus E (GPa)               & 8.7                          & 12.1                         & 18.5                         \\ \hline
Shear modulus   G (GPa)                 & 3.3                          & 4.5                          & 7.3                          \\ \hline
Poisson's   Ratio ${\rm \nu}$                     & 0.30                         & 0.33                         & 0.30                         \\ \hline
Pugh's Ratio   B/G                      & 2.2                          & 2.7                          & 1.9                          \\ \hline
Vickers   Hardness ${\rm H_V}$ (GPa)    & 1.6                          & 1.5                          & 3.1                          \\ \hline
Yield   Strength ${\rm \sigma_y}$ (GPa) & 0.5                          & 0.5                          & 1.0                          \\ \hline
\multicolumn{4}{|c|}{\multirow{4}{*}{\textit{\textbf{Universal Elastic Anisotropy Index}}}}                                          \\
\multicolumn{4}{|c|}{}                                                                                                               \\
\multicolumn{4}{|c|}{}                                                                                                               \\
\multicolumn{4}{|c|}{}                                                                                                               \\ \hline
${\rm A^U}$                             & 0.8                          & 1.1                          & 0.7                          \\ \hline
\end{tabular}
\end{table}

\section{Conclusion}
We have investigated the mechanical property, elastic modulus of 2D, quasi-2D and 3D hybrid organic-inorganic lead iodide perovskite thin films experimentally and substantiated these results through DFT simulations. We found that the elastic modulus of 2D, as well as quasi-2D Ruddlesden Popper perovskites, are lower than their 3D counterparts. This lower value of elastic modulus for Ruddlesden Popper perovskites is due to the presence of soft long chain organic layers between the inorganic octahedra and the weak Van Der Waals interactions between theses organic layers. These results demonstrate that variations in the subunits, the number of inorganic octahedra in between the long organic chain molecules of these 2D perovskites, have a huge effect on their mechanical properties. The mechanical properties of perovskite thus mainly depends on the soft organic layers, the weak Van Der Waals interface between these organic layers, and the stiff inorganic layers. The more the number of inorganic layers (n) present in between the long organic layers the more the value of the elastic modulus. Also, the preferential alignment of inorganic octahedra parallel to the substrate in the case of 2D perovskite further adds to its lower out-of-plane elastic modulus value.  The study of the out-of-plane mechanical properties of 2D, quasi-2D and 3D perovskite thin films along with their observed anisotropy can significantly enhance the understanding for designing perovskite based stable systems for applications like flexible and wearable electronics. The study of mechanical behaviour of these layered perovskite materials can significantly enhance the understanding of interaction between two subsequent inorganic (or organic) slabs, which can further be correlated with other material properties like anisotropic charge and phonon transport. These studies can provide valuable insights into the structure-property relationship in these materials~\cite{ElBallouli2020}, allowing detailed design of devices with desired material (mechanical and electronic) properties. \\

\section{Conflicts of Interest}
There are no conflicts of interest to declare.

\section{Acknowledgement}
Y.G. acknowledges support from the Institute Postdoctoral Fellowship, IIT Bombay. We acknowledge experimental support from the National Center for Photovoltaic Research and Education (NCPRE) at IIT Bombay.

\bibliography{ref}

\begin{thebibliography}{42}%
\makeatletter
\providecommand \@ifxundefined [1]{%
 \@ifx{#1\undefined}
}%
\providecommand \@ifnum [1]{%
 \ifnum #1\expandafter \@firstoftwo
 \else \expandafter \@secondoftwo
 \fi
}%
\providecommand \@ifx [1]{%
 \ifx #1\expandafter \@firstoftwo
 \else \expandafter \@secondoftwo
 \fi
}%
\providecommand \natexlab [1]{#1}%
\providecommand \enquote  [1]{``#1''}%
\providecommand \bibnamefont  [1]{#1}%
\providecommand \bibfnamefont [1]{#1}%
\providecommand \citenamefont [1]{#1}%
\providecommand \href@noop [0]{\@secondoftwo}%
\providecommand \href [0]{\begingroup \@sanitize@url \@href}%
\providecommand \@href[1]{\@@startlink{#1}\@@href}%
\providecommand \@@href[1]{\endgroup#1\@@endlink}%
\providecommand \@sanitize@url [0]{\catcode `\\12\catcode `\$12\catcode
  `\&12\catcode `\#12\catcode `\^12\catcode `\_12\catcode `\%12\relax}%
\providecommand \@@startlink[1]{}%
\providecommand \@@endlink[0]{}%
\providecommand \url  [0]{\begingroup\@sanitize@url \@url }%
\providecommand \@url [1]{\endgroup\@href {#1}{\urlprefix }}%
\providecommand \urlprefix  [0]{URL }%
\providecommand \Eprint [0]{\href }%
\providecommand \doibase [0]{https://doi.org/}%
\providecommand \selectlanguage [0]{\@gobble}%
\providecommand \bibinfo  [0]{\@secondoftwo}%
\providecommand \bibfield  [0]{\@secondoftwo}%
\providecommand \translation [1]{[#1]}%
\providecommand \BibitemOpen [0]{}%
\providecommand \bibitemStop [0]{}%
\providecommand \bibitemNoStop [0]{.\EOS\space}%
\providecommand \EOS [0]{\spacefactor3000\relax}%
\providecommand \BibitemShut  [1]{\csname bibitem#1\endcsname}%
\let\auto@bib@innerbib\@empty
%</preamble>
\bibitem [{\citenamefont {Maes}\ \emph {et~al.}(2018)\citenamefont {Maes},
  \citenamefont {Balcaen}, \citenamefont {Drijvers}, \citenamefont {Zhao},
  \citenamefont {Roo}, \citenamefont {Vantomme}, \citenamefont {Vanhaecke},
  \citenamefont {Geiregat},\ and\ \citenamefont {Hens}}]{Maes2018}%
  \BibitemOpen
  \bibfield  {author} {\bibinfo {author} {\bibfnamefont {J.}~\bibnamefont
  {Maes}}, \bibinfo {author} {\bibfnamefont {L.}~\bibnamefont {Balcaen}},
  \bibinfo {author} {\bibfnamefont {E.}~\bibnamefont {Drijvers}}, \bibinfo
  {author} {\bibfnamefont {Q.}~\bibnamefont {Zhao}}, \bibinfo {author}
  {\bibfnamefont {J.~D.}\ \bibnamefont {Roo}}, \bibinfo {author} {\bibfnamefont
  {A.}~\bibnamefont {Vantomme}}, \bibinfo {author} {\bibfnamefont
  {F.}~\bibnamefont {Vanhaecke}}, \bibinfo {author} {\bibfnamefont
  {P.}~\bibnamefont {Geiregat}},\ and\ \bibinfo {author} {\bibfnamefont
  {Z.}~\bibnamefont {Hens}},\ }\bibfield  {title} {\bibinfo {title} {Light
  absorption coefficient of {CsPbBr}3 perovskite nanocrystals},\ }\href
  {https://doi.org/10.1021/acs.jpclett.8b01065} {\bibfield  {journal} {\bibinfo
   {journal} {The Journal of Physical Chemistry Letters}\ }\textbf {\bibinfo
  {volume} {9}},\ \bibinfo {pages} {3093} (\bibinfo {year} {2018})}\BibitemShut
  {NoStop}%
\bibitem [{\citenamefont {Park}(2015)}]{Park2015}%
  \BibitemOpen
  \bibfield  {author} {\bibinfo {author} {\bibfnamefont {N.-G.}\ \bibnamefont
  {Park}},\ }\bibfield  {title} {\bibinfo {title} {Perovskite solar cells: an
  emerging photovoltaic technology},\ }\href
  {https://doi.org/10.1016/j.mattod.2014.07.007} {\bibfield  {journal}
  {\bibinfo  {journal} {Materials Today}\ }\textbf {\bibinfo {volume} {18}},\
  \bibinfo {pages} {65} (\bibinfo {year} {2015})}\BibitemShut {NoStop}%
\bibitem [{\citenamefont {Ruf}\ \emph {et~al.}(2019)\citenamefont {Ruf},
  \citenamefont {Ayg\"{u}ler}, \citenamefont {Giesbrecht}, \citenamefont
  {Rendenbach}, \citenamefont {Magin}, \citenamefont {Docampo}, \citenamefont
  {Kalt},\ and\ \citenamefont {Hetterich}}]{Ruf2019}%
  \BibitemOpen
  \bibfield  {author} {\bibinfo {author} {\bibfnamefont {F.}~\bibnamefont
  {Ruf}}, \bibinfo {author} {\bibfnamefont {M.~F.}\ \bibnamefont
  {Ayg\"{u}ler}}, \bibinfo {author} {\bibfnamefont {N.}~\bibnamefont
  {Giesbrecht}}, \bibinfo {author} {\bibfnamefont {B.}~\bibnamefont
  {Rendenbach}}, \bibinfo {author} {\bibfnamefont {A.}~\bibnamefont {Magin}},
  \bibinfo {author} {\bibfnamefont {P.}~\bibnamefont {Docampo}}, \bibinfo
  {author} {\bibfnamefont {H.}~\bibnamefont {Kalt}},\ and\ \bibinfo {author}
  {\bibfnamefont {M.}~\bibnamefont {Hetterich}},\ }\bibfield  {title} {\bibinfo
  {title} {Temperature-dependent studies of exciton binding energy and
  phase-transition suppression in (cs, {FA}, {MA})pb(i, br)3 perovskites},\
  }\href {https://doi.org/10.1063/1.5083792} {\bibfield  {journal} {\bibinfo
  {journal} {{APL} Materials}\ }\textbf {\bibinfo {volume} {7}},\ \bibinfo
  {pages} {031113} (\bibinfo {year} {2019})}\BibitemShut {NoStop}%
\bibitem [{\citenamefont {Wang}\ \emph {et~al.}(2016)\citenamefont {Wang},
  \citenamefont {Yuan}, \citenamefont {Duan}, \citenamefont {Huang},
  \citenamefont {Wei}, \citenamefont {Liu}, \citenamefont {Wang},\ and\
  \citenamefont {Li}}]{Wang2016}%
  \BibitemOpen
  \bibfield  {author} {\bibinfo {author} {\bibfnamefont {L.}~\bibnamefont
  {Wang}}, \bibinfo {author} {\bibfnamefont {G.~D.}\ \bibnamefont {Yuan}},
  \bibinfo {author} {\bibfnamefont {R.~F.}\ \bibnamefont {Duan}}, \bibinfo
  {author} {\bibfnamefont {F.}~\bibnamefont {Huang}}, \bibinfo {author}
  {\bibfnamefont {T.~B.}\ \bibnamefont {Wei}}, \bibinfo {author} {\bibfnamefont
  {Z.~Q.}\ \bibnamefont {Liu}}, \bibinfo {author} {\bibfnamefont {J.~X.}\
  \bibnamefont {Wang}},\ and\ \bibinfo {author} {\bibfnamefont {J.~M.}\
  \bibnamefont {Li}},\ }\bibfield  {title} {\bibinfo {title} {Tunable bandgap
  in hybrid perovskite {CH}3nh3pb(br3-{yXy}) single crystals and photodetector
  applications},\ }\href {https://doi.org/10.1063/1.4948312} {\bibfield
  {journal} {\bibinfo  {journal} {{AIP} Advances}\ }\textbf {\bibinfo {volume}
  {6}},\ \bibinfo {pages} {045115} (\bibinfo {year} {2016})}\BibitemShut
  {NoStop}%
\bibitem [{\citenamefont {Dong}\ \emph {et~al.}(2017)\citenamefont {Dong},
  \citenamefont {Deng}, \citenamefont {Hu}, \citenamefont {Song}, \citenamefont
  {Qiao}, \citenamefont {Yang}, \citenamefont {Zhang}, \citenamefont {Cai},
  \citenamefont {Tang},\ and\ \citenamefont {Song}}]{Dong2017}%
  \BibitemOpen
  \bibfield  {author} {\bibinfo {author} {\bibfnamefont {D.}~\bibnamefont
  {Dong}}, \bibinfo {author} {\bibfnamefont {H.}~\bibnamefont {Deng}}, \bibinfo
  {author} {\bibfnamefont {C.}~\bibnamefont {Hu}}, \bibinfo {author}
  {\bibfnamefont {H.}~\bibnamefont {Song}}, \bibinfo {author} {\bibfnamefont
  {K.}~\bibnamefont {Qiao}}, \bibinfo {author} {\bibfnamefont {X.}~\bibnamefont
  {Yang}}, \bibinfo {author} {\bibfnamefont {J.}~\bibnamefont {Zhang}},
  \bibinfo {author} {\bibfnamefont {F.}~\bibnamefont {Cai}}, \bibinfo {author}
  {\bibfnamefont {J.}~\bibnamefont {Tang}},\ and\ \bibinfo {author}
  {\bibfnamefont {H.}~\bibnamefont {Song}},\ }\bibfield  {title} {\bibinfo
  {title} {Bandgap tunable csx({CH}3nh3)1-{xPbI}3perovskite nanowires by
  aqueous solution synthesis for optoelectronic devices},\ }\href
  {https://doi.org/10.1039/c6nr06636d} {\bibfield  {journal} {\bibinfo
  {journal} {Nanoscale}\ }\textbf {\bibinfo {volume} {9}},\ \bibinfo {pages}
  {1567} (\bibinfo {year} {2017})}\BibitemShut {NoStop}%
\bibitem [{\citenamefont {Herz}(2017)}]{Herz2017}%
  \BibitemOpen
  \bibfield  {author} {\bibinfo {author} {\bibfnamefont {L.~M.}\ \bibnamefont
  {Herz}},\ }\bibfield  {title} {\bibinfo {title} {Charge-carrier mobilities in
  metal halide perovskites: Fundamental mechanisms and limits},\ }\href
  {https://doi.org/10.1021/acsenergylett.7b00276} {\bibfield  {journal}
  {\bibinfo  {journal} {{ACS} Energy Letters}\ }\textbf {\bibinfo {volume}
  {2}},\ \bibinfo {pages} {1539} (\bibinfo {year} {2017})}\BibitemShut
  {NoStop}%
\bibitem [{\citenamefont {Dittrich}\ \emph {et~al.}(2016)\citenamefont
  {Dittrich}, \citenamefont {Lang}, \citenamefont {Shargaieva}, \citenamefont
  {Rappich}, \citenamefont {Nickel}, \citenamefont {Unger},\ and\ \citenamefont
  {Rech}}]{Dittrich2016}%
  \BibitemOpen
  \bibfield  {author} {\bibinfo {author} {\bibfnamefont {T.}~\bibnamefont
  {Dittrich}}, \bibinfo {author} {\bibfnamefont {F.}~\bibnamefont {Lang}},
  \bibinfo {author} {\bibfnamefont {O.}~\bibnamefont {Shargaieva}}, \bibinfo
  {author} {\bibfnamefont {J.}~\bibnamefont {Rappich}}, \bibinfo {author}
  {\bibfnamefont {N.~H.}\ \bibnamefont {Nickel}}, \bibinfo {author}
  {\bibfnamefont {E.}~\bibnamefont {Unger}},\ and\ \bibinfo {author}
  {\bibfnamefont {B.}~\bibnamefont {Rech}},\ }\bibfield  {title} {\bibinfo
  {title} {Diffusion length of photo-generated charge carriers in layers and
  powders of {CH}3nh3pbi3 perovskite},\ }\href
  {https://doi.org/10.1063/1.4960641} {\bibfield  {journal} {\bibinfo
  {journal} {Applied Physics Letters}\ }\textbf {\bibinfo {volume} {109}},\
  \bibinfo {pages} {073901} (\bibinfo {year} {2016})}\BibitemShut {NoStop}%
\bibitem [{\citenamefont {Chen}\ \emph {et~al.}(2018)\citenamefont {Chen},
  \citenamefont {Sch\"{u}nemann}, \citenamefont {Song},\ and\ \citenamefont
  {T\"{u}ys\"{u}z}}]{Chen2018}%
  \BibitemOpen
  \bibfield  {author} {\bibinfo {author} {\bibfnamefont {K.}~\bibnamefont
  {Chen}}, \bibinfo {author} {\bibfnamefont {S.}~\bibnamefont
  {Sch\"{u}nemann}}, \bibinfo {author} {\bibfnamefont {S.}~\bibnamefont
  {Song}},\ and\ \bibinfo {author} {\bibfnamefont {H.}~\bibnamefont
  {T\"{u}ys\"{u}z}},\ }\bibfield  {title} {\bibinfo {title} {Structural effects
  on optoelectronic properties of halide perovskites},\ }\href
  {https://doi.org/10.1039/c8cs00212f} {\bibfield  {journal} {\bibinfo
  {journal} {Chemical Society Reviews}\ }\textbf {\bibinfo {volume} {47}},\
  \bibinfo {pages} {7045} (\bibinfo {year} {2018})}\BibitemShut {NoStop}%
\bibitem [{\citenamefont {Tu}\ \emph {et~al.}(2018{\natexlab{a}})\citenamefont
  {Tu}, \citenamefont {Spanopoulos}, \citenamefont {Yasaei}, \citenamefont
  {Stoumpos}, \citenamefont {Kanatzidis}, \citenamefont {Shekhawat},\ and\
  \citenamefont {Dravid}}]{Tu2018_i}%
  \BibitemOpen
  \bibfield  {author} {\bibinfo {author} {\bibfnamefont {Q.}~\bibnamefont
  {Tu}}, \bibinfo {author} {\bibfnamefont {I.}~\bibnamefont {Spanopoulos}},
  \bibinfo {author} {\bibfnamefont {P.}~\bibnamefont {Yasaei}}, \bibinfo
  {author} {\bibfnamefont {C.~C.}\ \bibnamefont {Stoumpos}}, \bibinfo {author}
  {\bibfnamefont {M.~G.}\ \bibnamefont {Kanatzidis}}, \bibinfo {author}
  {\bibfnamefont {G.~S.}\ \bibnamefont {Shekhawat}},\ and\ \bibinfo {author}
  {\bibfnamefont {V.~P.}\ \bibnamefont {Dravid}},\ }\bibfield  {title}
  {\bibinfo {title} {Stretching and breaking of ultrathin 2d hybrid
  organic{\textendash}inorganic perovskites},\ }\href
  {https://doi.org/10.1021/acsnano.8b05623} {\bibfield  {journal} {\bibinfo
  {journal} {{ACS} Nano}\ }\textbf {\bibinfo {volume} {12}},\ \bibinfo {pages}
  {10347} (\bibinfo {year} {2018}{\natexlab{a}})}\BibitemShut {NoStop}%
\bibitem [{\citenamefont {Tu}\ \emph {et~al.}(2018{\natexlab{b}})\citenamefont
  {Tu}, \citenamefont {Spanopoulos}, \citenamefont {Hao}, \citenamefont
  {Wolverton}, \citenamefont {Kanatzidis}, \citenamefont {Shekhawat},\ and\
  \citenamefont {Dravid}}]{Tu2018}%
  \BibitemOpen
  \bibfield  {author} {\bibinfo {author} {\bibfnamefont {Q.}~\bibnamefont
  {Tu}}, \bibinfo {author} {\bibfnamefont {I.}~\bibnamefont {Spanopoulos}},
  \bibinfo {author} {\bibfnamefont {S.}~\bibnamefont {Hao}}, \bibinfo {author}
  {\bibfnamefont {C.}~\bibnamefont {Wolverton}}, \bibinfo {author}
  {\bibfnamefont {M.~G.}\ \bibnamefont {Kanatzidis}}, \bibinfo {author}
  {\bibfnamefont {G.~S.}\ \bibnamefont {Shekhawat}},\ and\ \bibinfo {author}
  {\bibfnamefont {V.~P.}\ \bibnamefont {Dravid}},\ }\bibfield  {title}
  {\bibinfo {title} {Out-of-plane mechanical properties of 2d hybrid
  organic{\textendash}inorganic perovskites by nanoindentation},\ }\href
  {https://doi.org/10.1021/acsami.8b05138} {\bibfield  {journal} {\bibinfo
  {journal} {{ACS} Applied Materials {\&} Interfaces}\ }\textbf {\bibinfo
  {volume} {10}},\ \bibinfo {pages} {22167} (\bibinfo {year}
  {2018}{\natexlab{b}})}\BibitemShut {NoStop}%
\bibitem [{\citenamefont {Spanopoulos}\ \emph {et~al.}(2019)\citenamefont
  {Spanopoulos}, \citenamefont {Hadar}, \citenamefont {Ke}, \citenamefont {Tu},
  \citenamefont {Chen}, \citenamefont {Tsai}, \citenamefont {He}, \citenamefont
  {Shekhawat}, \citenamefont {Dravid}, \citenamefont {Wasielewski},
  \citenamefont {Mohite}, \citenamefont {Stoumpos},\ and\ \citenamefont
  {Kanatzidis}}]{Spanopoulos2019}%
  \BibitemOpen
  \bibfield  {author} {\bibinfo {author} {\bibfnamefont {I.}~\bibnamefont
  {Spanopoulos}}, \bibinfo {author} {\bibfnamefont {I.}~\bibnamefont {Hadar}},
  \bibinfo {author} {\bibfnamefont {W.}~\bibnamefont {Ke}}, \bibinfo {author}
  {\bibfnamefont {Q.}~\bibnamefont {Tu}}, \bibinfo {author} {\bibfnamefont
  {M.}~\bibnamefont {Chen}}, \bibinfo {author} {\bibfnamefont {H.}~\bibnamefont
  {Tsai}}, \bibinfo {author} {\bibfnamefont {Y.}~\bibnamefont {He}}, \bibinfo
  {author} {\bibfnamefont {G.}~\bibnamefont {Shekhawat}}, \bibinfo {author}
  {\bibfnamefont {V.~P.}\ \bibnamefont {Dravid}}, \bibinfo {author}
  {\bibfnamefont {M.~R.}\ \bibnamefont {Wasielewski}}, \bibinfo {author}
  {\bibfnamefont {A.~D.}\ \bibnamefont {Mohite}}, \bibinfo {author}
  {\bibfnamefont {C.~C.}\ \bibnamefont {Stoumpos}},\ and\ \bibinfo {author}
  {\bibfnamefont {M.~G.}\ \bibnamefont {Kanatzidis}},\ }\bibfield  {title}
  {\bibinfo {title} {Uniaxial expansion of the 2d ruddlesden{\textendash}popper
  perovskite family for improved environmental stability},\ }\href
  {https://doi.org/10.1021/jacs.9b01327} {\bibfield  {journal} {\bibinfo
  {journal} {Journal of the American Chemical Society}\ }\textbf {\bibinfo
  {volume} {141}},\ \bibinfo {pages} {5518} (\bibinfo {year}
  {2019})}\BibitemShut {NoStop}%
\bibitem [{\citenamefont {Reyes-Martinez}\ \emph {et~al.}(2020)\citenamefont
  {Reyes-Martinez}, \citenamefont {Tan}, \citenamefont {Kakekhani},
  \citenamefont {Banerjee}, \citenamefont {Zhumekenov}, \citenamefont {Peng},
  \citenamefont {Bakr}, \citenamefont {Rappe},\ and\ \citenamefont
  {Loo}}]{ReyesMartinez2020}%
  \BibitemOpen
  \bibfield  {author} {\bibinfo {author} {\bibfnamefont {M.~A.}\ \bibnamefont
  {Reyes-Martinez}}, \bibinfo {author} {\bibfnamefont {P.}~\bibnamefont {Tan}},
  \bibinfo {author} {\bibfnamefont {A.}~\bibnamefont {Kakekhani}}, \bibinfo
  {author} {\bibfnamefont {S.}~\bibnamefont {Banerjee}}, \bibinfo {author}
  {\bibfnamefont {A.~A.}\ \bibnamefont {Zhumekenov}}, \bibinfo {author}
  {\bibfnamefont {W.}~\bibnamefont {Peng}}, \bibinfo {author} {\bibfnamefont
  {O.~M.}\ \bibnamefont {Bakr}}, \bibinfo {author} {\bibfnamefont {A.~M.}\
  \bibnamefont {Rappe}},\ and\ \bibinfo {author} {\bibfnamefont {Y.-L.}\
  \bibnamefont {Loo}},\ }\bibfield  {title} {\bibinfo {title} {Unraveling the
  elastic properties of (quasi)two-dimensional hybrid perovskites: A joint
  experimental and theoretical study},\ }\href
  {https://doi.org/10.1021/acsami.0c02327} {\bibfield  {journal} {\bibinfo
  {journal} {{ACS} Applied Materials {\&} Interfaces}\ }\textbf {\bibinfo
  {volume} {12}},\ \bibinfo {pages} {17881} (\bibinfo {year}
  {2020})}\BibitemShut {NoStop}%
\bibitem [{\citenamefont {Li}\ \emph {et~al.}(2019)\citenamefont {Li},
  \citenamefont {Bi}, \citenamefont {Bu}, \citenamefont {Tang}, \citenamefont
  {Ouyang}, \citenamefont {Jiang},\ and\ \citenamefont {Song}}]{Li2019}%
  \BibitemOpen
  \bibfield  {author} {\bibinfo {author} {\bibfnamefont {Q.}~\bibnamefont
  {Li}}, \bibinfo {author} {\bibfnamefont {S.}~\bibnamefont {Bi}}, \bibinfo
  {author} {\bibfnamefont {J.}~\bibnamefont {Bu}}, \bibinfo {author}
  {\bibfnamefont {C.}~\bibnamefont {Tang}}, \bibinfo {author} {\bibfnamefont
  {Z.}~\bibnamefont {Ouyang}}, \bibinfo {author} {\bibfnamefont
  {C.}~\bibnamefont {Jiang}},\ and\ \bibinfo {author} {\bibfnamefont
  {J.}~\bibnamefont {Song}},\ }\bibfield  {title} {\bibinfo {title} {Atomic
  layer dependence of shear modulus in a two-dimensional single-crystal
  organic{\textendash}inorganic hybrid perovskite},\ }\href
  {https://doi.org/10.1021/acs.jpcc.9b02080} {\bibfield  {journal} {\bibinfo
  {journal} {The Journal of Physical Chemistry C}\ }\textbf {\bibinfo {volume}
  {123}},\ \bibinfo {pages} {15251} (\bibinfo {year} {2019})}\BibitemShut
  {NoStop}%
\bibitem [{\citenamefont {Gao}\ \emph {et~al.}(2020)\citenamefont {Gao},
  \citenamefont {Wei}, \citenamefont {Li}, \citenamefont {Tan},\ and\
  \citenamefont {Tang}}]{Gao2020}%
  \BibitemOpen
  \bibfield  {author} {\bibinfo {author} {\bibfnamefont {H.}~\bibnamefont
  {Gao}}, \bibinfo {author} {\bibfnamefont {W.}~\bibnamefont {Wei}}, \bibinfo
  {author} {\bibfnamefont {L.}~\bibnamefont {Li}}, \bibinfo {author}
  {\bibfnamefont {Y.}~\bibnamefont {Tan}},\ and\ \bibinfo {author}
  {\bibfnamefont {Y.}~\bibnamefont {Tang}},\ }\bibfield  {title} {\bibinfo
  {title} {Mechanical properties of a 2d lead-halide perovskite,
  (c6h5ch2nh3)2pbcl4, by nanoindentation and first-principles calculations},\
  }\href {https://doi.org/10.1021/acs.jpcc.0c04283} {\bibfield  {journal}
  {\bibinfo  {journal} {The Journal of Physical Chemistry C}\ }\textbf
  {\bibinfo {volume} {124}},\ \bibinfo {pages} {19204} (\bibinfo {year}
  {2020})}\BibitemShut {NoStop}%
\bibitem [{\citenamefont {Rathore}\ \emph {et~al.}(2021)\citenamefont
  {Rathore}, \citenamefont {Han}, \citenamefont {Kumar}, \citenamefont
  {Leong},\ and\ \citenamefont {Singh}}]{Rathore2021}%
  \BibitemOpen
  \bibfield  {author} {\bibinfo {author} {\bibfnamefont {S.}~\bibnamefont
  {Rathore}}, \bibinfo {author} {\bibfnamefont {G.}~\bibnamefont {Han}},
  \bibinfo {author} {\bibfnamefont {A.}~\bibnamefont {Kumar}}, \bibinfo
  {author} {\bibfnamefont {W.~L.}\ \bibnamefont {Leong}},\ and\ \bibinfo
  {author} {\bibfnamefont {A.}~\bibnamefont {Singh}},\ }\bibfield  {title}
  {\bibinfo {title} {Elastic modulus tailoring in {CH}3nh3pbi3 perovskite
  system by the introduction of two dimensionality using (5-{AVA})2pbi4},\
  }\href {https://doi.org/10.1016/j.solener.2021.05.027} {\bibfield  {journal}
  {\bibinfo  {journal} {Solar Energy}\ }\textbf {\bibinfo {volume} {224}},\
  \bibinfo {pages} {27} (\bibinfo {year} {2021})}\BibitemShut {NoStop}%
\bibitem [{\citenamefont {Oliver}\ and\ \citenamefont
  {Pharr}(1992)}]{Oliver1992}%
  \BibitemOpen
  \bibfield  {author} {\bibinfo {author} {\bibfnamefont {W.}~\bibnamefont
  {Oliver}}\ and\ \bibinfo {author} {\bibfnamefont {G.}~\bibnamefont {Pharr}},\
  }\bibfield  {title} {\bibinfo {title} {An improved technique for determining
  hardness and elastic modulus using load and displacement sensing indentation
  experiments},\ }\href {https://doi.org/10.1557/jmr.1992.1564} {\bibfield
  {journal} {\bibinfo  {journal} {Journal of Materials Research}\ }\textbf
  {\bibinfo {volume} {7}},\ \bibinfo {pages} {1564} (\bibinfo {year}
  {1992})}\BibitemShut {NoStop}%
\bibitem [{\citenamefont {Pharr}\ and\ \citenamefont
  {Oliver}(1992)}]{Pharr1992}%
  \BibitemOpen
  \bibfield  {author} {\bibinfo {author} {\bibfnamefont {G.}~\bibnamefont
  {Pharr}}\ and\ \bibinfo {author} {\bibfnamefont {W.}~\bibnamefont {Oliver}},\
  }\bibfield  {title} {\bibinfo {title} {Measurement of thin film mechanical
  properties using nanoindentation},\ }\href
  {https://doi.org/10.1557/s0883769400041634} {\bibfield  {journal} {\bibinfo
  {journal} {{MRS} Bulletin}\ }\textbf {\bibinfo {volume} {17}},\ \bibinfo
  {pages} {28} (\bibinfo {year} {1992})}\BibitemShut {NoStop}%
\bibitem [{\citenamefont {Oliver}\ and\ \citenamefont
  {Pharr}(2004)}]{Oliver2004}%
  \BibitemOpen
  \bibfield  {author} {\bibinfo {author} {\bibfnamefont {W.}~\bibnamefont
  {Oliver}}\ and\ \bibinfo {author} {\bibfnamefont {G.}~\bibnamefont {Pharr}},\
  }\bibfield  {title} {\bibinfo {title} {Measurement of hardness and elastic
  modulus by instrumented indentation: Advances in understanding and
  refinements to methodology},\ }\href
  {https://doi.org/10.1557/jmr.2004.19.1.3} {\bibfield  {journal} {\bibinfo
  {journal} {Journal of Materials Research}\ }\textbf {\bibinfo {volume}
  {19}},\ \bibinfo {pages} {3} (\bibinfo {year} {2004})}\BibitemShut {NoStop}%
\bibitem [{\citenamefont {Sun}\ \emph {et~al.}(2015)\citenamefont {Sun},
  \citenamefont {Fang}, \citenamefont {Kieslich}, \citenamefont {White},\ and\
  \citenamefont {Cheetham}}]{Sun2015}%
  \BibitemOpen
  \bibfield  {author} {\bibinfo {author} {\bibfnamefont {S.}~\bibnamefont
  {Sun}}, \bibinfo {author} {\bibfnamefont {Y.}~\bibnamefont {Fang}}, \bibinfo
  {author} {\bibfnamefont {G.}~\bibnamefont {Kieslich}}, \bibinfo {author}
  {\bibfnamefont {T.~J.}\ \bibnamefont {White}},\ and\ \bibinfo {author}
  {\bibfnamefont {A.~K.}\ \bibnamefont {Cheetham}},\ }\bibfield  {title}
  {\bibinfo {title} {Mechanical properties of organic - inorganic halide
  perovskites, {CH}3nh3pbx3(x = i, br and cl), by nanoindentation},\ }\href
  {https://doi.org/10.1039/c5ta03331d} {\bibfield  {journal} {\bibinfo
  {journal} {Journal of Materials Chemistry A}\ }\textbf {\bibinfo {volume}
  {3}},\ \bibinfo {pages} {18450} (\bibinfo {year} {2015})}\BibitemShut
  {NoStop}%
\bibitem [{\citenamefont {Feng}(2014)}]{Feng2014}%
  \BibitemOpen
  \bibfield  {author} {\bibinfo {author} {\bibfnamefont {J.}~\bibnamefont
  {Feng}},\ }\bibfield  {title} {\bibinfo {title} {Mechanical properties of
  hybrid organic-inorganic {CH}3nh3bx3 (b = sn, pb; x = br, i) perovskites for
  solar cell absorbers},\ }\href {https://doi.org/10.1063/1.4885256} {\bibfield
   {journal} {\bibinfo  {journal} {{APL} Materials}\ }\textbf {\bibinfo
  {volume} {2}},\ \bibinfo {pages} {081801} (\bibinfo {year}
  {2014})}\BibitemShut {NoStop}%
\bibitem [{\citenamefont {Rakita}\ \emph {et~al.}(2015)\citenamefont {Rakita},
  \citenamefont {Cohen}, \citenamefont {Kedem}, \citenamefont {Hodes},\ and\
  \citenamefont {Cahen}}]{Rakita2015}%
  \BibitemOpen
  \bibfield  {author} {\bibinfo {author} {\bibfnamefont {Y.}~\bibnamefont
  {Rakita}}, \bibinfo {author} {\bibfnamefont {S.~R.}\ \bibnamefont {Cohen}},
  \bibinfo {author} {\bibfnamefont {N.~K.}\ \bibnamefont {Kedem}}, \bibinfo
  {author} {\bibfnamefont {G.}~\bibnamefont {Hodes}},\ and\ \bibinfo {author}
  {\bibfnamefont {D.}~\bibnamefont {Cahen}},\ }\bibfield  {title} {\bibinfo
  {title} {Mechanical properties of {APbX}3 (a = cs or {CH}3nh3; x = i or br)
  perovskite single crystals},\ }\href {https://doi.org/10.1557/mrc.2015.69}
  {\bibfield  {journal} {\bibinfo  {journal} {{MRS} Communications}\ }\textbf
  {\bibinfo {volume} {5}},\ \bibinfo {pages} {623} (\bibinfo {year}
  {2015})}\BibitemShut {NoStop}%
\bibitem [{\citenamefont {Corso}(2016)}]{DalCorso2016}%
  \BibitemOpen
  \bibfield  {author} {\bibinfo {author} {\bibfnamefont {A.~D.}\ \bibnamefont
  {Corso}},\ }\bibfield  {title} {\bibinfo {title} {Elastic constants of
  beryllium: a first-principles investigation},\ }\href
  {https://doi.org/10.1088/0953-8984/28/7/075401} {\bibfield  {journal}
  {\bibinfo  {journal} {Journal of Physics: Condensed Matter}\ }\textbf
  {\bibinfo {volume} {28}},\ \bibinfo {pages} {075401} (\bibinfo {year}
  {2016})}\BibitemShut {NoStop}%
\bibitem [{\citenamefont {Giannozzi}\ \emph {et~al.}(2009)\citenamefont
  {Giannozzi}, \citenamefont {Baroni}, \citenamefont {Bonini}, \citenamefont
  {Calandra}, \citenamefont {Car}, \citenamefont {Cavazzoni}, \citenamefont
  {Ceresoli}, \citenamefont {Chiarotti}, \citenamefont {Cococcioni},
  \citenamefont {Dabo}, \citenamefont {Corso}, \citenamefont {de~Gironcoli},
  \citenamefont {Fabris}, \citenamefont {Fratesi}, \citenamefont {Gebauer},
  \citenamefont {Gerstmann}, \citenamefont {Gougoussis}, \citenamefont
  {Kokalj}, \citenamefont {Lazzeri}, \citenamefont {Martin-Samos},
  \citenamefont {Marzari}, \citenamefont {Mauri}, \citenamefont {Mazzarello},
  \citenamefont {Paolini}, \citenamefont {Pasquarello}, \citenamefont
  {Paulatto}, \citenamefont {Sbraccia}, \citenamefont {Scandolo}, \citenamefont
  {Sclauzero}, \citenamefont {Seitsonen}, \citenamefont {Smogunov},
  \citenamefont {Umari},\ and\ \citenamefont {Wentzcovitch}}]{Giannozzi2009}%
  \BibitemOpen
  \bibfield  {author} {\bibinfo {author} {\bibfnamefont {P.}~\bibnamefont
  {Giannozzi}}, \bibinfo {author} {\bibfnamefont {S.}~\bibnamefont {Baroni}},
  \bibinfo {author} {\bibfnamefont {N.}~\bibnamefont {Bonini}}, \bibinfo
  {author} {\bibfnamefont {M.}~\bibnamefont {Calandra}}, \bibinfo {author}
  {\bibfnamefont {R.}~\bibnamefont {Car}}, \bibinfo {author} {\bibfnamefont
  {C.}~\bibnamefont {Cavazzoni}}, \bibinfo {author} {\bibfnamefont
  {D.}~\bibnamefont {Ceresoli}}, \bibinfo {author} {\bibfnamefont {G.~L.}\
  \bibnamefont {Chiarotti}}, \bibinfo {author} {\bibfnamefont {M.}~\bibnamefont
  {Cococcioni}}, \bibinfo {author} {\bibfnamefont {I.}~\bibnamefont {Dabo}},
  \bibinfo {author} {\bibfnamefont {A.~D.}\ \bibnamefont {Corso}}, \bibinfo
  {author} {\bibfnamefont {S.}~\bibnamefont {de~Gironcoli}}, \bibinfo {author}
  {\bibfnamefont {S.}~\bibnamefont {Fabris}}, \bibinfo {author} {\bibfnamefont
  {G.}~\bibnamefont {Fratesi}}, \bibinfo {author} {\bibfnamefont
  {R.}~\bibnamefont {Gebauer}}, \bibinfo {author} {\bibfnamefont
  {U.}~\bibnamefont {Gerstmann}}, \bibinfo {author} {\bibfnamefont
  {C.}~\bibnamefont {Gougoussis}}, \bibinfo {author} {\bibfnamefont
  {A.}~\bibnamefont {Kokalj}}, \bibinfo {author} {\bibfnamefont
  {M.}~\bibnamefont {Lazzeri}}, \bibinfo {author} {\bibfnamefont
  {L.}~\bibnamefont {Martin-Samos}}, \bibinfo {author} {\bibfnamefont
  {N.}~\bibnamefont {Marzari}}, \bibinfo {author} {\bibfnamefont
  {F.}~\bibnamefont {Mauri}}, \bibinfo {author} {\bibfnamefont
  {R.}~\bibnamefont {Mazzarello}}, \bibinfo {author} {\bibfnamefont
  {S.}~\bibnamefont {Paolini}}, \bibinfo {author} {\bibfnamefont
  {A.}~\bibnamefont {Pasquarello}}, \bibinfo {author} {\bibfnamefont
  {L.}~\bibnamefont {Paulatto}}, \bibinfo {author} {\bibfnamefont
  {C.}~\bibnamefont {Sbraccia}}, \bibinfo {author} {\bibfnamefont
  {S.}~\bibnamefont {Scandolo}}, \bibinfo {author} {\bibfnamefont
  {G.}~\bibnamefont {Sclauzero}}, \bibinfo {author} {\bibfnamefont {A.~P.}\
  \bibnamefont {Seitsonen}}, \bibinfo {author} {\bibfnamefont {A.}~\bibnamefont
  {Smogunov}}, \bibinfo {author} {\bibfnamefont {P.}~\bibnamefont {Umari}},\
  and\ \bibinfo {author} {\bibfnamefont {R.~M.}\ \bibnamefont {Wentzcovitch}},\
  }\bibfield  {title} {\bibinfo {title} {{QUANTUM} {ESPRESSO}: a modular and
  open-source software project for quantum simulations of materials},\ }\href
  {https://doi.org/10.1088/0953-8984/21/39/395502} {\bibfield  {journal}
  {\bibinfo  {journal} {Journal of Physics: Condensed Matter}\ }\textbf
  {\bibinfo {volume} {21}},\ \bibinfo {pages} {395502} (\bibinfo {year}
  {2009})}\BibitemShut {NoStop}%
\bibitem [{\citenamefont {Giannozzi}\ \emph {et~al.}(2017)\citenamefont
  {Giannozzi}, \citenamefont {Andreussi}, \citenamefont {Brumme}, \citenamefont
  {Bunau}, \citenamefont {Nardelli}, \citenamefont {Calandra}, \citenamefont
  {Car}, \citenamefont {Cavazzoni}, \citenamefont {Ceresoli}, \citenamefont
  {Cococcioni}, \citenamefont {Colonna}, \citenamefont {Carnimeo},
  \citenamefont {Corso}, \citenamefont {de~Gironcoli}, \citenamefont {Delugas},
  \citenamefont {DiStasio}, \citenamefont {Ferretti}, \citenamefont {Floris},
  \citenamefont {Fratesi}, \citenamefont {Fugallo}, \citenamefont {Gebauer},
  \citenamefont {Gerstmann}, \citenamefont {Giustino}, \citenamefont {Gorni},
  \citenamefont {Jia}, \citenamefont {Kawamura}, \citenamefont {Ko},
  \citenamefont {Kokalj}, \citenamefont {K\"{u}{\c{c}}\"{u}kbenli},
  \citenamefont {Lazzeri}, \citenamefont {Marsili}, \citenamefont {Marzari},
  \citenamefont {Mauri}, \citenamefont {Nguyen}, \citenamefont {Nguyen},
  \citenamefont {de-la Roza}, \citenamefont {Paulatto}, \citenamefont
  {Ponc{\'{e}}}, \citenamefont {Rocca}, \citenamefont {Sabatini}, \citenamefont
  {Santra}, \citenamefont {Schlipf}, \citenamefont {Seitsonen}, \citenamefont
  {Smogunov}, \citenamefont {Timrov}, \citenamefont {Thonhauser}, \citenamefont
  {Umari}, \citenamefont {Vast}, \citenamefont {Wu},\ and\ \citenamefont
  {Baroni}}]{Giannozzi2017}%
  \BibitemOpen
  \bibfield  {author} {\bibinfo {author} {\bibfnamefont {P.}~\bibnamefont
  {Giannozzi}}, \bibinfo {author} {\bibfnamefont {O.}~\bibnamefont
  {Andreussi}}, \bibinfo {author} {\bibfnamefont {T.}~\bibnamefont {Brumme}},
  \bibinfo {author} {\bibfnamefont {O.}~\bibnamefont {Bunau}}, \bibinfo
  {author} {\bibfnamefont {M.~B.}\ \bibnamefont {Nardelli}}, \bibinfo {author}
  {\bibfnamefont {M.}~\bibnamefont {Calandra}}, \bibinfo {author}
  {\bibfnamefont {R.}~\bibnamefont {Car}}, \bibinfo {author} {\bibfnamefont
  {C.}~\bibnamefont {Cavazzoni}}, \bibinfo {author} {\bibfnamefont
  {D.}~\bibnamefont {Ceresoli}}, \bibinfo {author} {\bibfnamefont
  {M.}~\bibnamefont {Cococcioni}}, \bibinfo {author} {\bibfnamefont
  {N.}~\bibnamefont {Colonna}}, \bibinfo {author} {\bibfnamefont
  {I.}~\bibnamefont {Carnimeo}}, \bibinfo {author} {\bibfnamefont {A.~D.}\
  \bibnamefont {Corso}}, \bibinfo {author} {\bibfnamefont {S.}~\bibnamefont
  {de~Gironcoli}}, \bibinfo {author} {\bibfnamefont {P.}~\bibnamefont
  {Delugas}}, \bibinfo {author} {\bibfnamefont {R.~A.}\ \bibnamefont
  {DiStasio}}, \bibinfo {author} {\bibfnamefont {A.}~\bibnamefont {Ferretti}},
  \bibinfo {author} {\bibfnamefont {A.}~\bibnamefont {Floris}}, \bibinfo
  {author} {\bibfnamefont {G.}~\bibnamefont {Fratesi}}, \bibinfo {author}
  {\bibfnamefont {G.}~\bibnamefont {Fugallo}}, \bibinfo {author} {\bibfnamefont
  {R.}~\bibnamefont {Gebauer}}, \bibinfo {author} {\bibfnamefont
  {U.}~\bibnamefont {Gerstmann}}, \bibinfo {author} {\bibfnamefont
  {F.}~\bibnamefont {Giustino}}, \bibinfo {author} {\bibfnamefont
  {T.}~\bibnamefont {Gorni}}, \bibinfo {author} {\bibfnamefont
  {J.}~\bibnamefont {Jia}}, \bibinfo {author} {\bibfnamefont {M.}~\bibnamefont
  {Kawamura}}, \bibinfo {author} {\bibfnamefont {H.-Y.}\ \bibnamefont {Ko}},
  \bibinfo {author} {\bibfnamefont {A.}~\bibnamefont {Kokalj}}, \bibinfo
  {author} {\bibfnamefont {E.}~\bibnamefont {K\"{u}{\c{c}}\"{u}kbenli}},
  \bibinfo {author} {\bibfnamefont {M.}~\bibnamefont {Lazzeri}}, \bibinfo
  {author} {\bibfnamefont {M.}~\bibnamefont {Marsili}}, \bibinfo {author}
  {\bibfnamefont {N.}~\bibnamefont {Marzari}}, \bibinfo {author} {\bibfnamefont
  {F.}~\bibnamefont {Mauri}}, \bibinfo {author} {\bibfnamefont {N.~L.}\
  \bibnamefont {Nguyen}}, \bibinfo {author} {\bibfnamefont {H.-V.}\
  \bibnamefont {Nguyen}}, \bibinfo {author} {\bibfnamefont {A.~O.}\
  \bibnamefont {de-la Roza}}, \bibinfo {author} {\bibfnamefont
  {L.}~\bibnamefont {Paulatto}}, \bibinfo {author} {\bibfnamefont
  {S.}~\bibnamefont {Ponc{\'{e}}}}, \bibinfo {author} {\bibfnamefont
  {D.}~\bibnamefont {Rocca}}, \bibinfo {author} {\bibfnamefont
  {R.}~\bibnamefont {Sabatini}}, \bibinfo {author} {\bibfnamefont
  {B.}~\bibnamefont {Santra}}, \bibinfo {author} {\bibfnamefont
  {M.}~\bibnamefont {Schlipf}}, \bibinfo {author} {\bibfnamefont {A.~P.}\
  \bibnamefont {Seitsonen}}, \bibinfo {author} {\bibfnamefont {A.}~\bibnamefont
  {Smogunov}}, \bibinfo {author} {\bibfnamefont {I.}~\bibnamefont {Timrov}},
  \bibinfo {author} {\bibfnamefont {T.}~\bibnamefont {Thonhauser}}, \bibinfo
  {author} {\bibfnamefont {P.}~\bibnamefont {Umari}}, \bibinfo {author}
  {\bibfnamefont {N.}~\bibnamefont {Vast}}, \bibinfo {author} {\bibfnamefont
  {X.}~\bibnamefont {Wu}},\ and\ \bibinfo {author} {\bibfnamefont
  {S.}~\bibnamefont {Baroni}},\ }\bibfield  {title} {\bibinfo {title} {Advanced
  capabilities for materials modelling with quantum {ESPRESSO}},\ }\href
  {https://doi.org/10.1088/1361-648x/aa8f79} {\bibfield  {journal} {\bibinfo
  {journal} {Journal of Physics: Condensed Matter}\ }\textbf {\bibinfo {volume}
  {29}},\ \bibinfo {pages} {465901} (\bibinfo {year} {2017})}\BibitemShut
  {NoStop}%
\bibitem [{\citenamefont {Perdew}\ \emph {et~al.}(1998)\citenamefont {Perdew},
  \citenamefont {Burke},\ and\ \citenamefont {Ernzerhof}}]{Perdew1998}%
  \BibitemOpen
  \bibfield  {author} {\bibinfo {author} {\bibfnamefont {J.~P.}\ \bibnamefont
  {Perdew}}, \bibinfo {author} {\bibfnamefont {K.}~\bibnamefont {Burke}},\ and\
  \bibinfo {author} {\bibfnamefont {M.}~\bibnamefont {Ernzerhof}},\ }\bibfield
  {title} {\bibinfo {title} {Perdew, burke, and ernzerhof reply:},\ }\href
  {https://doi.org/10.1103/physrevlett.80.891} {\bibfield  {journal} {\bibinfo
  {journal} {Physical Review Letters}\ }\textbf {\bibinfo {volume} {80}},\
  \bibinfo {pages} {891} (\bibinfo {year} {1998})}\BibitemShut {NoStop}%
\bibitem [{\citenamefont {Pugh}(1954)}]{Pugh1954}%
  \BibitemOpen
  \bibfield  {author} {\bibinfo {author} {\bibfnamefont {S.}~\bibnamefont
  {Pugh}},\ }\bibfield  {title} {\bibinfo {title} {{XCII}. relations between
  the elastic moduli and the plastic properties of polycrystalline pure
  metals},\ }\href {https://doi.org/10.1080/14786440808520496} {\bibfield
  {journal} {\bibinfo  {journal} {The London, Edinburgh, and Dublin
  Philosophical Magazine and Journal of Science}\ }\textbf {\bibinfo {volume}
  {45}},\ \bibinfo {pages} {823} (\bibinfo {year} {1954})}\BibitemShut
  {NoStop}%
\bibitem [{\citenamefont {Chen}\ \emph {et~al.}(2011)\citenamefont {Chen},
  \citenamefont {Niu}, \citenamefont {Li},\ and\ \citenamefont
  {Li}}]{Chen2011}%
  \BibitemOpen
  \bibfield  {author} {\bibinfo {author} {\bibfnamefont {X.-Q.}\ \bibnamefont
  {Chen}}, \bibinfo {author} {\bibfnamefont {H.}~\bibnamefont {Niu}}, \bibinfo
  {author} {\bibfnamefont {D.}~\bibnamefont {Li}},\ and\ \bibinfo {author}
  {\bibfnamefont {Y.}~\bibnamefont {Li}},\ }\bibfield  {title} {\bibinfo
  {title} {Modeling hardness of polycrystalline materials and bulk metallic
  glasses},\ }\href {https://doi.org/10.1016/j.intermet.2011.03.026} {\bibfield
   {journal} {\bibinfo  {journal} {Intermetallics}\ }\textbf {\bibinfo {volume}
  {19}},\ \bibinfo {pages} {1275} (\bibinfo {year} {2011})}\BibitemShut
  {NoStop}%
\bibitem [{\citenamefont {Zaddach}\ \emph {et~al.}(2013)\citenamefont
  {Zaddach}, \citenamefont {Niu}, \citenamefont {Koch},\ and\ \citenamefont
  {Irving}}]{Zaddach2013}%
  \BibitemOpen
  \bibfield  {author} {\bibinfo {author} {\bibfnamefont {A.~J.}\ \bibnamefont
  {Zaddach}}, \bibinfo {author} {\bibfnamefont {C.}~\bibnamefont {Niu}},
  \bibinfo {author} {\bibfnamefont {C.~C.}\ \bibnamefont {Koch}},\ and\
  \bibinfo {author} {\bibfnamefont {D.~L.}\ \bibnamefont {Irving}},\ }\bibfield
   {title} {\bibinfo {title} {Mechanical properties and stacking fault energies
  of {NiFeCrCoMn} high-entropy alloy},\ }\href
  {https://doi.org/10.1007/s11837-013-0771-4} {\bibfield  {journal} {\bibinfo
  {journal} {{JOM}}\ }\textbf {\bibinfo {volume} {65}},\ \bibinfo {pages}
  {1780} (\bibinfo {year} {2013})}\BibitemShut {NoStop}%
\bibitem [{\citenamefont {Ranganathan}\ and\ \citenamefont
  {Ostoja-Starzewski}(2008)}]{Ranganathan2008}%
  \BibitemOpen
  \bibfield  {author} {\bibinfo {author} {\bibfnamefont {S.~I.}\ \bibnamefont
  {Ranganathan}}\ and\ \bibinfo {author} {\bibfnamefont {M.}~\bibnamefont
  {Ostoja-Starzewski}},\ }\bibfield  {title} {\bibinfo {title} {Universal
  elastic anisotropy index},\ }\bibfield  {journal} {\bibinfo  {journal}
  {Physical Review Letters}\ }\textbf {\bibinfo {volume} {101}},\ \href
  {https://doi.org/10.1103/physrevlett.101.055504}
  {10.1103/physrevlett.101.055504} (\bibinfo {year} {2008})\BibitemShut
  {NoStop}%
\bibitem [{\citenamefont {Sch\"{o}tz}\ \emph {et~al.}(2020)\citenamefont
  {Sch\"{o}tz}, \citenamefont {Askar}, \citenamefont {Peng}, \citenamefont
  {Seeberger}, \citenamefont {Gujar}, \citenamefont {Thelakkat}, \citenamefont
  {K\"{o}hler}, \citenamefont {Huettner}, \citenamefont {Bakr}, \citenamefont
  {Shankar},\ and\ \citenamefont {Panzer}}]{Schtz2020}%
  \BibitemOpen
  \bibfield  {author} {\bibinfo {author} {\bibfnamefont {K.}~\bibnamefont
  {Sch\"{o}tz}}, \bibinfo {author} {\bibfnamefont {A.~M.}\ \bibnamefont
  {Askar}}, \bibinfo {author} {\bibfnamefont {W.}~\bibnamefont {Peng}},
  \bibinfo {author} {\bibfnamefont {D.}~\bibnamefont {Seeberger}}, \bibinfo
  {author} {\bibfnamefont {T.~P.}\ \bibnamefont {Gujar}}, \bibinfo {author}
  {\bibfnamefont {M.}~\bibnamefont {Thelakkat}}, \bibinfo {author}
  {\bibfnamefont {A.}~\bibnamefont {K\"{o}hler}}, \bibinfo {author}
  {\bibfnamefont {S.}~\bibnamefont {Huettner}}, \bibinfo {author}
  {\bibfnamefont {O.~M.}\ \bibnamefont {Bakr}}, \bibinfo {author}
  {\bibfnamefont {K.}~\bibnamefont {Shankar}},\ and\ \bibinfo {author}
  {\bibfnamefont {F.}~\bibnamefont {Panzer}},\ }\bibfield  {title} {\bibinfo
  {title} {Double peak emission in lead halide perovskites by
  self-absorption},\ }\href {https://doi.org/10.1039/c9tc06251c} {\bibfield
  {journal} {\bibinfo  {journal} {Journal of Materials Chemistry C}\ }\textbf
  {\bibinfo {volume} {8}},\ \bibinfo {pages} {2289} (\bibinfo {year}
  {2020})}\BibitemShut {NoStop}%
\bibitem [{\citenamefont {Hwang}\ and\ \citenamefont {Lee}(2018)}]{Hwang2018}%
  \BibitemOpen
  \bibfield  {author} {\bibinfo {author} {\bibfnamefont {B.}~\bibnamefont
  {Hwang}}\ and\ \bibinfo {author} {\bibfnamefont {J.-S.}\ \bibnamefont
  {Lee}},\ }\bibfield  {title} {\bibinfo {title} {2d perovskite-based
  self-aligned lateral heterostructure photodetectors utilizing vapor
  deposition},\ }\href {https://doi.org/10.1002/adom.201801356} {\bibfield
  {journal} {\bibinfo  {journal} {Advanced Optical Materials}\ }\textbf
  {\bibinfo {volume} {7}},\ \bibinfo {pages} {1801356} (\bibinfo {year}
  {2018})}\BibitemShut {NoStop}%
\bibitem [{\citenamefont {Blancon}\ \emph {et~al.}(2017)\citenamefont
  {Blancon}, \citenamefont {Tsai}, \citenamefont {Nie}, \citenamefont
  {Stoumpos}, \citenamefont {Pedesseau}, \citenamefont {Katan}, \citenamefont
  {Kepenekian}, \citenamefont {Soe}, \citenamefont {Appavoo}, \citenamefont
  {Sfeir}, \citenamefont {Tretiak}, \citenamefont {Ajayan}, \citenamefont
  {Kanatzidis}, \citenamefont {Even}, \citenamefont {Crochet},\ and\
  \citenamefont {Mohite}}]{Blancon2017}%
  \BibitemOpen
  \bibfield  {author} {\bibinfo {author} {\bibfnamefont {J.-C.}\ \bibnamefont
  {Blancon}}, \bibinfo {author} {\bibfnamefont {H.}~\bibnamefont {Tsai}},
  \bibinfo {author} {\bibfnamefont {W.}~\bibnamefont {Nie}}, \bibinfo {author}
  {\bibfnamefont {C.~C.}\ \bibnamefont {Stoumpos}}, \bibinfo {author}
  {\bibfnamefont {L.}~\bibnamefont {Pedesseau}}, \bibinfo {author}
  {\bibfnamefont {C.}~\bibnamefont {Katan}}, \bibinfo {author} {\bibfnamefont
  {M.}~\bibnamefont {Kepenekian}}, \bibinfo {author} {\bibfnamefont {C.~M.~M.}\
  \bibnamefont {Soe}}, \bibinfo {author} {\bibfnamefont {K.}~\bibnamefont
  {Appavoo}}, \bibinfo {author} {\bibfnamefont {M.~Y.}\ \bibnamefont {Sfeir}},
  \bibinfo {author} {\bibfnamefont {S.}~\bibnamefont {Tretiak}}, \bibinfo
  {author} {\bibfnamefont {P.~M.}\ \bibnamefont {Ajayan}}, \bibinfo {author}
  {\bibfnamefont {M.~G.}\ \bibnamefont {Kanatzidis}}, \bibinfo {author}
  {\bibfnamefont {J.}~\bibnamefont {Even}}, \bibinfo {author} {\bibfnamefont
  {J.~J.}\ \bibnamefont {Crochet}},\ and\ \bibinfo {author} {\bibfnamefont
  {A.~D.}\ \bibnamefont {Mohite}},\ }\bibfield  {title} {\bibinfo {title}
  {Extremely efficient internal exciton dissociation through edge states in
  layered 2d perovskites},\ }\href {https://doi.org/10.1126/science.aal4211}
  {\bibfield  {journal} {\bibinfo  {journal} {Science}\ }\textbf {\bibinfo
  {volume} {355}},\ \bibinfo {pages} {1288} (\bibinfo {year}
  {2017})}\BibitemShut {NoStop}%
\bibitem [{\citenamefont {Cao}\ \emph {et~al.}(2015)\citenamefont {Cao},
  \citenamefont {Stoumpos}, \citenamefont {Farha}, \citenamefont {Hupp},\ and\
  \citenamefont {Kanatzidis}}]{Cao2015}%
  \BibitemOpen
  \bibfield  {author} {\bibinfo {author} {\bibfnamefont {D.~H.}\ \bibnamefont
  {Cao}}, \bibinfo {author} {\bibfnamefont {C.~C.}\ \bibnamefont {Stoumpos}},
  \bibinfo {author} {\bibfnamefont {O.~K.}\ \bibnamefont {Farha}}, \bibinfo
  {author} {\bibfnamefont {J.~T.}\ \bibnamefont {Hupp}},\ and\ \bibinfo
  {author} {\bibfnamefont {M.~G.}\ \bibnamefont {Kanatzidis}},\ }\bibfield
  {title} {\bibinfo {title} {2d homologous perovskites as light-absorbing
  materials for solar cell applications},\ }\href
  {https://doi.org/10.1021/jacs.5b03796} {\bibfield  {journal} {\bibinfo
  {journal} {Journal of the American Chemical Society}\ }\textbf {\bibinfo
  {volume} {137}},\ \bibinfo {pages} {7843} (\bibinfo {year}
  {2015})}\BibitemShut {NoStop}%
\bibitem [{\citenamefont {Soe}\ \emph {et~al.}(2017)\citenamefont {Soe},
  \citenamefont {Nie}, \citenamefont {Stoumpos}, \citenamefont {Tsai},
  \citenamefont {Blancon}, \citenamefont {Liu}, \citenamefont {Even},
  \citenamefont {Marks}, \citenamefont {Mohite},\ and\ \citenamefont
  {Kanatzidis}}]{Soe2017}%
  \BibitemOpen
  \bibfield  {author} {\bibinfo {author} {\bibfnamefont {C.~M.~M.}\
  \bibnamefont {Soe}}, \bibinfo {author} {\bibfnamefont {W.}~\bibnamefont
  {Nie}}, \bibinfo {author} {\bibfnamefont {C.~C.}\ \bibnamefont {Stoumpos}},
  \bibinfo {author} {\bibfnamefont {H.}~\bibnamefont {Tsai}}, \bibinfo {author}
  {\bibfnamefont {J.-C.}\ \bibnamefont {Blancon}}, \bibinfo {author}
  {\bibfnamefont {F.}~\bibnamefont {Liu}}, \bibinfo {author} {\bibfnamefont
  {J.}~\bibnamefont {Even}}, \bibinfo {author} {\bibfnamefont {T.~J.}\
  \bibnamefont {Marks}}, \bibinfo {author} {\bibfnamefont {A.~D.}\ \bibnamefont
  {Mohite}},\ and\ \bibinfo {author} {\bibfnamefont {M.~G.}\ \bibnamefont
  {Kanatzidis}},\ }\bibfield  {title} {\bibinfo {title} {Understanding film
  formation morphology and orientation in high member 2d ruddlesden-popper
  perovskites for high-efficiency solar cells},\ }\href
  {https://doi.org/10.1002/aenm.201700979} {\bibfield  {journal} {\bibinfo
  {journal} {Advanced Energy Materials}\ }\textbf {\bibinfo {volume} {8}},\
  \bibinfo {pages} {1700979} (\bibinfo {year} {2017})}\BibitemShut {NoStop}%
\bibitem [{\citenamefont {Venkatesan}\ \emph {et~al.}(2018)\citenamefont
  {Venkatesan}, \citenamefont {Labram},\ and\ \citenamefont
  {Chabinyc}}]{Venkatesan2018}%
  \BibitemOpen
  \bibfield  {author} {\bibinfo {author} {\bibfnamefont {N.~R.}\ \bibnamefont
  {Venkatesan}}, \bibinfo {author} {\bibfnamefont {J.~G.}\ \bibnamefont
  {Labram}},\ and\ \bibinfo {author} {\bibfnamefont {M.~L.}\ \bibnamefont
  {Chabinyc}},\ }\bibfield  {title} {\bibinfo {title} {Charge-carrier dynamics
  and crystalline texture of layered ruddlesden - popper hybrid lead iodide
  perovskite thin films},\ }\href
  {https://doi.org/10.1021/acsenergylett.7b01245} {\bibfield  {journal}
  {\bibinfo  {journal} {{ACS} Energy Letters}\ }\textbf {\bibinfo {volume}
  {3}},\ \bibinfo {pages} {380} (\bibinfo {year} {2018})}\BibitemShut {NoStop}%
\bibitem [{\citenamefont {Qiu}\ \emph {et~al.}(2018)\citenamefont {Qiu},
  \citenamefont {Zheng}, \citenamefont {Xia}, \citenamefont {Chao},
  \citenamefont {Chen},\ and\ \citenamefont {Huang}}]{Qiu2018}%
  \BibitemOpen
  \bibfield  {author} {\bibinfo {author} {\bibfnamefont {J.}~\bibnamefont
  {Qiu}}, \bibinfo {author} {\bibfnamefont {Y.}~\bibnamefont {Zheng}}, \bibinfo
  {author} {\bibfnamefont {Y.}~\bibnamefont {Xia}}, \bibinfo {author}
  {\bibfnamefont {L.}~\bibnamefont {Chao}}, \bibinfo {author} {\bibfnamefont
  {Y.}~\bibnamefont {Chen}},\ and\ \bibinfo {author} {\bibfnamefont
  {W.}~\bibnamefont {Huang}},\ }\bibfield  {title} {\bibinfo {title} {Rapid
  crystallization for efficient 2d ruddlesden-popper (2drp) perovskite solar
  cells},\ }\href {https://doi.org/10.1002/adfm.201806831} {\bibfield
  {journal} {\bibinfo  {journal} {Advanced Functional Materials}\ }\textbf
  {\bibinfo {volume} {29}},\ \bibinfo {pages} {1806831} (\bibinfo {year}
  {2018})}\BibitemShut {NoStop}%
\bibitem [{\citenamefont {Fan}\ \emph {et~al.}(2016)\citenamefont {Fan},
  \citenamefont {Gu}, \citenamefont {Liang}, \citenamefont {Luo}, \citenamefont
  {Chen}, \citenamefont {Zheng},\ and\ \citenamefont {Zhang}}]{fan2016}%
  \BibitemOpen
  \bibfield  {author} {\bibinfo {author} {\bibfnamefont {P.}~\bibnamefont
  {Fan}}, \bibinfo {author} {\bibfnamefont {D.}~\bibnamefont {Gu}}, \bibinfo
  {author} {\bibfnamefont {G.-X.}\ \bibnamefont {Liang}}, \bibinfo {author}
  {\bibfnamefont {J.-T.}\ \bibnamefont {Luo}}, \bibinfo {author} {\bibfnamefont
  {J.-L.}\ \bibnamefont {Chen}}, \bibinfo {author} {\bibfnamefont {Z.-H.}\
  \bibnamefont {Zheng}},\ and\ \bibinfo {author} {\bibfnamefont {D.-P.}\
  \bibnamefont {Zhang}},\ }\bibfield  {title} {\bibinfo {title}
  {High-performance perovskite ch3nh3pbi3 thin films for solar cells prepared
  by single-source physical vapour deposition},\ }\href
  {https://doi.org/10.1038/srep29910} {\bibfield  {journal} {\bibinfo
  {journal} {Scientific Reports}\ }\textbf {\bibinfo {volume} {6}},\ \bibinfo
  {pages} {7843} (\bibinfo {year} {2016})}\BibitemShut {NoStop}%
\bibitem [{\citenamefont {Sun}\ \emph {et~al.}(2017)\citenamefont {Sun},
  \citenamefont {Isikgor}, \citenamefont {Deng}, \citenamefont {Wei},
  \citenamefont {Kieslich}, \citenamefont {Bristowe}, \citenamefont {Ouyang},\
  and\ \citenamefont {Cheetham}}]{Sun2017}%
  \BibitemOpen
  \bibfield  {author} {\bibinfo {author} {\bibfnamefont {S.}~\bibnamefont
  {Sun}}, \bibinfo {author} {\bibfnamefont {F.~H.}\ \bibnamefont {Isikgor}},
  \bibinfo {author} {\bibfnamefont {Z.}~\bibnamefont {Deng}}, \bibinfo {author}
  {\bibfnamefont {F.}~\bibnamefont {Wei}}, \bibinfo {author} {\bibfnamefont
  {G.}~\bibnamefont {Kieslich}}, \bibinfo {author} {\bibfnamefont {P.~D.}\
  \bibnamefont {Bristowe}}, \bibinfo {author} {\bibfnamefont {J.}~\bibnamefont
  {Ouyang}},\ and\ \bibinfo {author} {\bibfnamefont {A.~K.}\ \bibnamefont
  {Cheetham}},\ }\bibfield  {title} {\bibinfo {title} {Factors influencing the
  mechanical properties of formamidinium lead halides and related hybrid
  perovskites},\ }\href {https://doi.org/10.1002/cssc.201701762} {\bibfield
  {journal} {\bibinfo  {journal} {{ChemSusChem}}\ }\textbf {\bibinfo {volume}
  {10}},\ \bibinfo {pages} {3683} (\bibinfo {year} {2017})}\BibitemShut
  {NoStop}%
\bibitem [{\citenamefont {Li}\ \emph {et~al.}(2014)\citenamefont {Li},
  \citenamefont {Thirumurugan}, \citenamefont {Barton}, \citenamefont {Lin},
  \citenamefont {Henke}, \citenamefont {Yeung}, \citenamefont {Wharmby},
  \citenamefont {Bithell}, \citenamefont {Howard},\ and\ \citenamefont
  {Cheetham}}]{Li2014}%
  \BibitemOpen
  \bibfield  {author} {\bibinfo {author} {\bibfnamefont {W.}~\bibnamefont
  {Li}}, \bibinfo {author} {\bibfnamefont {A.}~\bibnamefont {Thirumurugan}},
  \bibinfo {author} {\bibfnamefont {P.~T.}\ \bibnamefont {Barton}}, \bibinfo
  {author} {\bibfnamefont {Z.}~\bibnamefont {Lin}}, \bibinfo {author}
  {\bibfnamefont {S.}~\bibnamefont {Henke}}, \bibinfo {author} {\bibfnamefont
  {H.~H.-M.}\ \bibnamefont {Yeung}}, \bibinfo {author} {\bibfnamefont {M.~T.}\
  \bibnamefont {Wharmby}}, \bibinfo {author} {\bibfnamefont {E.~G.}\
  \bibnamefont {Bithell}}, \bibinfo {author} {\bibfnamefont {C.~J.}\
  \bibnamefont {Howard}},\ and\ \bibinfo {author} {\bibfnamefont {A.~K.}\
  \bibnamefont {Cheetham}},\ }\bibfield  {title} {\bibinfo {title} {Mechanical
  tunability via hydrogen bonding in metal - organic frameworks with the
  perovskite architecture},\ }\href {https://doi.org/10.1021/ja500618z}
  {\bibfield  {journal} {\bibinfo  {journal} {Journal of the American Chemical
  Society}\ }\textbf {\bibinfo {volume} {136}},\ \bibinfo {pages} {7801}
  (\bibinfo {year} {2014})}\BibitemShut {NoStop}%
\bibitem [{\citenamefont {Tan}\ \emph {et~al.}(2012)\citenamefont {Tan},
  \citenamefont {Jain},\ and\ \citenamefont {Cheetham}}]{Tan2012}%
  \BibitemOpen
  \bibfield  {author} {\bibinfo {author} {\bibfnamefont {J.-C.}\ \bibnamefont
  {Tan}}, \bibinfo {author} {\bibfnamefont {P.}~\bibnamefont {Jain}},\ and\
  \bibinfo {author} {\bibfnamefont {A.~K.}\ \bibnamefont {Cheetham}},\
  }\bibfield  {title} {\bibinfo {title} {Influence of ligand field
  stabilization energy on the elastic properties of multiferroic {MOFs} with
  the perovskite architecture},\ }\href {https://doi.org/10.1039/c2dt12300b}
  {\bibfield  {journal} {\bibinfo  {journal} {Dalton Transactions}\ }\textbf
  {\bibinfo {volume} {41}},\ \bibinfo {pages} {3949} (\bibinfo {year}
  {2012})}\BibitemShut {NoStop}%
\bibitem [{\citenamefont {Tu}\ \emph {et~al.}(2020)\citenamefont {Tu},
  \citenamefont {Spanopoulos}, \citenamefont {Vasileiadou}, \citenamefont {Li},
  \citenamefont {Kanatzidis}, \citenamefont {Shekhawat},\ and\ \citenamefont
  {Dravid}}]{Tu2020}%
  \BibitemOpen
  \bibfield  {author} {\bibinfo {author} {\bibfnamefont {Q.}~\bibnamefont
  {Tu}}, \bibinfo {author} {\bibfnamefont {I.}~\bibnamefont {Spanopoulos}},
  \bibinfo {author} {\bibfnamefont {E.~S.}\ \bibnamefont {Vasileiadou}},
  \bibinfo {author} {\bibfnamefont {X.}~\bibnamefont {Li}}, \bibinfo {author}
  {\bibfnamefont {M.~G.}\ \bibnamefont {Kanatzidis}}, \bibinfo {author}
  {\bibfnamefont {G.~S.}\ \bibnamefont {Shekhawat}},\ and\ \bibinfo {author}
  {\bibfnamefont {V.~P.}\ \bibnamefont {Dravid}},\ }\bibfield  {title}
  {\bibinfo {title} {Exploring the factors affecting the mechanical properties
  of 2d hybrid organic{\textendash}inorganic perovskites},\ }\href
  {https://doi.org/10.1021/acsami.0c02313} {\bibfield  {journal} {\bibinfo
  {journal} {{ACS} Applied Materials {\&} Interfaces}\ }\textbf {\bibinfo
  {volume} {12}},\ \bibinfo {pages} {20440} (\bibinfo {year}
  {2020})}\BibitemShut {NoStop}%
\bibitem [{\citenamefont {El-Ballouli}\ \emph {et~al.}(2020)\citenamefont
  {El-Ballouli}, \citenamefont {Bakr},\ and\ \citenamefont
  {Mohammed}}]{ElBallouli2020}%
  \BibitemOpen
  \bibfield  {author} {\bibinfo {author} {\bibfnamefont {A.~O.}\ \bibnamefont
  {El-Ballouli}}, \bibinfo {author} {\bibfnamefont {O.~M.}\ \bibnamefont
  {Bakr}},\ and\ \bibinfo {author} {\bibfnamefont {O.~F.}\ \bibnamefont
  {Mohammed}},\ }\bibfield  {title} {\bibinfo {title} {Structurally tunable
  two-dimensional layered perovskites: From confinement and enhanced charge
  transport to prolonged hot carrier cooling dynamics},\ }\href
  {https://doi.org/10.1021/acs.jpclett.0c00359} {\bibfield  {journal} {\bibinfo
   {journal} {The Journal of Physical Chemistry Letters}\ }\textbf {\bibinfo
  {volume} {11}},\ \bibinfo {pages} {5705} (\bibinfo {year}
  {2020})}\BibitemShut {NoStop}%
\end{thebibliography}%

\end{document}